\documentclass[fleqn,usenatbib]{mnras}

\usepackage{newtxtext,newtxmath}
\usepackage[T1]{fontenc}

\DeclareRobustCommand{\VAN}[3]{#2}
\let\VANthebibliography\thebibliography
\def\thebibliography{\DeclareRobustCommand{\VAN}[3]{##3}\VANthebibliography}

\usepackage{graphicx}	% Including figure files
\usepackage{amsmath}	% Advanced maths commands

\usepackage[flushleft]{threeparttable}
\usepackage{hyperref}
\usepackage{tabularx}
\usepackage{xcolor}
\usepackage{booktabs}

\usepackage{adjustbox}
\usepackage[flushleft]{threeparttable}
\usepackage{hyperref}
\usepackage{tabularx}
\usepackage{xcolor}
\usepackage{booktabs}
\usepackage{graphicx}	% Including figure files
\usepackage{amsmath}	% Advanced maths commands
\usepackage{multicol}        % Multi-column entries in tables
\usepackage{bm}		% Bold maths symbols, including upright Greek
\usepackage{pdflscape}	% Landscape pages
\usepackage[T1]{fontenc}
\usepackage{ae,aecompl}
\usepackage{newtxtext,newtxmath}
\usepackage[normalem]{ulem}

\def\Rn2 {\Romannum{2} }
\def\p {\textquotesingle}

\def\Rn2 {\Romannum{2} }

\title[Earliest optical observations of GRB 230328B]{Earliest simultaneous multi-color optical observations of GRB 230328B: from 41 seconds to the host-galaxy identification}

\author[Komesh et al.]{T. Komesh,$^{1,2,3}$\thanks{E-mail: toktarkhan.komesh@nu.edu.kz (TK)}
A. Pozanenko,$^{4,5}$\thanks{E-mail: apozanen@iki.rssi.ru (AP)}
N. Pankov,$^{4,5}$
A. Volnova,$^{4}$
P. Minaev,$^{4}$
R. Gill,$^{6,7}$
D. Berdikhan,$^{1}$\newauthor
B. Grossan,$^{1,8}$\thanks{Deceased.}
Z. Maksut,$^{1}$
Z. Abdullayev,$^{9,10}$
S. Belkin,$^{11}$
M. Krugov,$^{12}$
A. Moskvitin,$^{13}$
K. Baigarin,$^{1}$\newauthor
A. Tursynkan,$^{14}$
E. Klunko,$^{15}$
A. Tatarnikov,$^{16}$
S. Zheltoukhov,$^{16}$
V. Rumyantsev,$^{17}$
A. Volvach,$^{17}$
L. Volvach,$^{17}$\newauthor
O. A. Burkhonov,$^{18}$
S.A. Ehgamberdiev,$^{18,19}$
R. Inasaridze,$^{20,21}$
L. Elenin,$^{22}$
A. Krylov,$^{22}$
M. Zheltobryukhov,$^{23}$\newauthor
S. B. Pandey,$^{24}$
A. K. Ror,$^{24}$
R. Gupta,$^{25,26}$
V. Swain,$^{27}$
V. Bhalerao,$^{27}$
G. C. Anupama,$^{28}$
S. Barway,$^{28}$\newauthor
R. Sánchez-Ramírez,$^{29}$
A. J. Castro-Tirado,$^{29}$
S.~Antier,$^{30,31}$
P. Beniamini$^{32,7,33}$
and E. Abdikamalov$^{1,14}$
}

\date{Accepted 2026 August 20. Received 2026 August 5; in original form 2026 May 28}

\pubyear{\the\year{}}

\begin{document}
\label{firstpage}
\pagerange{\pageref{firstpage}--\pageref{lastpage}}
\maketitle

\begin{abstract}
We present a multiwavelength study of the long-duration gamma-ray burst
GRB~230328B, combining prompt $\gamma$-ray observations with exceptionally
early simultaneous optical photometry and extensive X-ray, radio, and
host-galaxy follow-up. NUTTELA-TAO began simultaneous $g'r'i'$ observations
only 41~s after the \textit{Swift}/BAT trigger, corresponding to approximately
16~s in the rest frame for the adopted host-galaxy photometric redshift of
$z_{\rm phot}=1.54\pm0.06$. Unlike sequential multiband measurements, these
observations provide an instantaneous optical spectral slope without
temporal-interpolation uncertainties. The early optical continuum remains statistically consistent with a
constant slope from approximately $10^{2}$ to $7\times10^{3}$~s, providing no
evidence for rapid dust destruction. The absence of significant color
evolution across the pronounced optical rebrightening at approximately
$4\times10^{3}$~s demonstrates that it is achromatic and favours a dynamical
rather than spectral origin. Broadband modelling shows that the optical,
X-ray, and radio evolution can be broadly explained by forward-shock emission
with late energy injection, although alternative scenarios cannot be excluded.
The red optical continuum is consistent with substantial line-of-sight
extinction in a massive, dusty host galaxy. Late-time observations are
insufficient to constrain a typical GRB-associated supernova at the adopted
redshift. GRB~230328B therefore provides a benchmark for connecting the
earliest simultaneous optical colors with afterglow dynamics and the host
environment of a representative long GRB.

\end{abstract}


%%%%%%%%%%%%%%%%%%%%%%%%%%%%%%%%%%%%%%%%%%%%%%%%%%

%%%%%%%%%%%%%%%%% BODY OF PAPER %%%%%%%%%%%%%%%%%%

\section{Introduction} 
\label{sec:intro}

Gamma-ray bursts (GRBs) are energetic explosions driven by relativistic jets. They are characterized by a \emph{prompt} phase, which emits $\gamma$-rays due to internal dissipation within the jet. This is followed by a broadband \emph{afterglow}, spanning X-rays to radio frequencies, generated as the jet interacts with the ambient medium (e.g., \citealt{Piran99Gamma}; \citealt{Kumar15Physics}). Beyond probing relativistic jet dynamics, particle acceleration \citep[e.g.,][]{Sironi11Particle}, and jet structure \citep{Beniamini19Observational, OConnor23structured}, GRBs also serve as powerful probes of star formation, metal enrichment, and the interstellar and intergalactic media across a wide redshift range \citep[e.g.,][]{Palmerio19Are, Gupta22Gamma, Graham23Surprising}. 
Despite decades of progress, key questions about the nature of GRBs remain, including the physical mechanism responsible for the prompt $\gamma$-ray emission, the composition and energy-dissipation processes of the relativistic jet, the structure and density profile of the circumburst environment, and the connection between GRB progenitors and their host-galaxy environments \citep[e.g.,][]{Kumar15Physics, GRB_review_2021AstL...47..791P}. Addressing these questions motivates coordinated multiwavelength observations from the earliest phases of the burst through the late afterglow and host-galaxy identification.

GRB durations exhibit a bimodal distribution  \citep{Kouveliotou93Identification}.Bursts with $T_{90} \leq 2$~s, where $T_{90}$ is the interval over which 90\% of the prompt $\gamma$-ray fluence is detected, are classified as short, while those with $T_{90} > 2$~s are classified as long \citep{Mazets81Catalog, Zhang25duration}. Since short and long GRBs tend to exhibit harder and softer spectra, they are commonly referred to as short-hard (Type~I) and long-soft (Type~II) bursts \citep{Zhang06Astrophysics, Zhang09Discerning}. This divide is largely tied to distinct progenitors. The short GRB~170817A, detected in coincidence with a gravitational wave signal and a subsequent kilonova, confirmed a binary neutron star merger origin for Type~I events \citep{Abbott17GW_GRB}. Long GRBs are associated with core-collapse supernovae (SNe)  (e.g., \citealt{Finneran25GRBSN}; \citealt{Woosley1993}). Examples include the associations of GRB~980425/SN~1998bw \citep{Galama98GRB_SN} and GRB~030329/SN~2003dh \citep{Hjorth03very}. However, this duration-based classification is not absolute \citep{Zhang25duration}. For example, GRB~211211A exhibited a duration of $T_{90} > 2$~s but showed clear kilonova signatures, indicating a merger origin despite its extended emission \citep{Troja22Nature}.

GRB afterglows are powerful tools for studying GRBs and their surroundings. The afterglow is governed by both intrinsic burst parameters \citep[e.g.,][]{Gao13complete} and the properties of the ambient medium \citep[][]{liang2013comprehensive, Tian22Constraining}. The afterglow signal can differentiate between a stellar wind profile and interstellar medium (ISM) \citep{Chevalier99Gamma}, and reveal transitions between the two, as seen in GRB~140423A \citep{Li20GRB140423A} and GRB~160625B \citep{Fraija17Theoretical}. \citet{Lazzati06clumpy_wind} utilized time-resolved spectroscopy to uncover a clumpy, extended wind environment around GRB~021004, highlighting the necessity of high-quality, wide-timescale multi-color data. GRBs act as bright ``backlights'' for their host galaxies \citep{Draine2002,Li2008}. Dust extinction in the ISM leaves a distinct imprint on the optical-UV spectrum of the burst \citep{Kann2006, schady2010dust, Melandri17Colour, Li18Large}. Several studies have reported color evolution from red to blue on short ($\sim 10^3$~s) timescales, consistent with dust destruction by the UV/X-ray radiation of the GRB (\citealt{Perna2003, Morgan2014}; \citealt{Komesh2023MN}; \citealt{Abdullayev2025}). However, single-band or non-simultaneous multiband observations introduce a degeneracy, as changes in the optical light curve may instead arise from intrinsic spectral evolution rather than dust destruction. Simultaneous multicolor photometry can help break this degeneracy by separating environmental effects from intrinsic emission physics \citep[for a comprehensive optical afterglow catalog, see][]{Dainotti2024MNRAS}.

While numerous early optical observations of GRB afterglows have been conducted (e.g., \citealt{Akerlof99Observation, Rykoff2009ApJ, Vestrand05Link, Klotz06Continuous, Gupta21GRB140102A, Xin23Prompt}; \citealt{Becerra2021}; \citealt{Becerra23Understanding, Cheng25GRB240825A}), simultaneous multicolor photometry remains rare. Robotic instruments such as RAPTOR, ROTSE, TORTORA, Pi of the Sky, and MASTER-net have captured optical flashes during or immediately following the $\gamma$-ray phase \citep[e.g.,][]{Vestrand05Link, Racusin2008Natur, Rykoff2009ApJ, Gorbovskoy12Prompt, Ror23Prompt, Gupta2026review}. These earliest optical signals are uniquely valuable because they are highly sensitive to the reverse shock propagating into the ejecta and the initial deceleration of the jet, providing crucial constraints on the jet's initial Lorentz factor and magnetization \citep[e.g.,][]{Sari99Predictions}. However, most detailed spectral tracking begins tens of minutes after the initial trigger \citep[e.g.,][]{Greiner2024, Kruhler2011, Gupta24Extremely}.

Late-time observations provide an essential complement to rapid GRB afterglow follow-up. Once the afterglow has faded sufficiently, deep imaging can isolate the underlying galaxy and establish the position of the transient relative to the host-galaxy light distribution. Accurate astrometry, combined with the angular separation and apparent magnitude of a candidate host, enables a quantitative estimate of the probability of chance coincidence and therefore a more robust GRB--host association \citep[e.g.,][]{Bloom2002}. This assessment is particularly important in crowded fields or complex environments containing neighbouring or interacting galaxies, where the nearest visible source may not necessarily be the true host. Late-time multiband imaging also provides host photometry with minimal afterglow contamination, allowing estimates of the photometric redshift, stellar mass, star-formation rate, age, and extinction when spectroscopy is unavailable. Furthermore, these observations establish the host-galaxy reference flux needed to search for an associated supernova. Observations extending from the early afterglow to the host-dominated phase are therefore necessary for connecting the properties of the transient with those of its galactic environment.

Recently, \citet{Komesh2023MN} reported the earliest known case of simultaneous \emph{g'r'i'} observations, beginning just 58 s after the Burst Alert Telescope (BAT) trigger for GRB~201015A. This was achieved using the Nazarbayev University Transient Telescope (NUTTELA-TAO) \citep{grossan2019emission, Maksut2021}, located at the Assy-Turgen Astrophysical Observatory \citep{Aimuratov25}, which can respond to alerts and point to targets within about 8 s. 

Using this capability, we present observations of the long-duration GRB~230328B. 
The scientific value of GRB~230328B lies in the combination of exceptionally early, strictly simultaneous multicolor photometry and extensive broadband follow-up. NUTTELA-TAO began simultaneous $g'r'i'$ observations only 41~s after the \textit{Swift}/BAT trigger, corresponding to approximately 16~s in the rest frame for the assumed photometric redshift of $z=1.54$. Unlike sequential multiband measurements, these observations provide an instantaneous optical spectral slope without requiring temporal interpolation across a rapidly varying light curve. The early optical continuum remains statistically consistent with a constant spectral slope from approximately $10^{2}$ to $7 \times 10^{3}$~s. This result provides no evidence for rapid dust destruction. Furthermore, the absence of significant color evolution across the pronounced optical rebrightening at approximately $4 \times 10^{3}$~s demonstrates that this feature is achromatic, favouring a dynamical origin, such as energy injection, rather than spectral evolution. Combined with the X-ray, radio, late-time optical and infrared observations, and host-galaxy characterization, these data make GRB~230328B a useful benchmark for connecting the earliest afterglow behaviour with broadband modelling and the dusty environment of an otherwise representative long GRB.

This paper is organized as follows. Section~\ref{sec:obs} describes the prompt $\gamma$-ray, optical, X-ray, radio, and late-time host-galaxy observations, together with the corresponding data-reduction procedures. Section~\ref{sec:res} presents the prompt-emission analysis, the temporal and spectral properties of the afterglow, the broadband spectral-energy-distribution fitting, and the physical modelling of the afterglow and its rebrightening. It also describes the identification and characterization of the host galaxy and the search for an associated supernova. Finally, Section~\ref{sec:conc} summarizes the principal results and their implications.

\section{Observations} 
\label{sec:obs}

\begin{figure*}
\centering
\includegraphics[width=\linewidth]{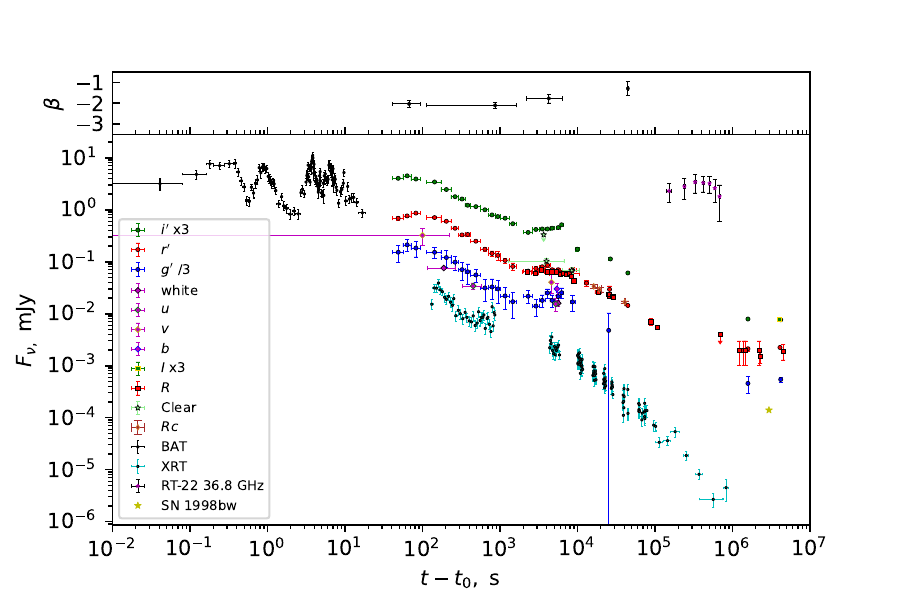}
\caption{Upper panel: the observed optical spectral slope $\beta$ (without correction for a host galaxy extinction) for GRB~230328B as a function of time. We determine the spectral slope at each epoch by fitting a power law to the $g'r'i'$ fluxes and plot the time-averaged results.
Bottom panel: multi-wavelength light curves including $\gamma$-ray, X-ray, optical, and radio data.
Here, $t$ denotes the midpoint time of each co-add and $t_0$ is the BAT trigger time, which occurred at 14:54:48 UT on 28 March 2023 \citep{Gropp2023GCN}. The large yellow star denotes the flux at the maximum of a light curve of SN 1998bw-like supernova as viewed from observer frame at the distance of the GRB 230328B source $z_{\rm phot}=1.54\pm0.06$. The parameters of the  light curve of SN 1998bw in observer frame was calculated using multicolor SN 1998bw light curve \citep{Clocchiatti2011} and proper transformation in the observer frame. }
\label{fig:LC1}
\end{figure*}

We define the \emph{early afterglow} phase as the first $\sim1\,$hour after the BAT trigger; all later times are referred to as the \emph{late afterglow}. We characterize the light curve flux by $F_{\nu} \propto t^{\alpha}\,\nu^{\beta}$. Distinct segments or features of the light curve are labeled with numerical subscripts. All magnitudes reported throughout the paper are in the AB system and have been corrected for Galactic extinction using the extinction maps of \citet{SF2011_Extinction}. The full light curve is shown in Fig.~\ref{fig:LC1}.

\subsection{Early afterglow observations}
\label{sec:obs_early}

GRB~230328B was triggered by the \textit{Swift}/BAT at 14:54:48 UT on 2023 March 28 \citep{Gropp2023GCN}.  The same \textit{Swift} GCN Circular reported an optical counterpart detected by \textit{Swift}/UVOT at R.A.~(J2000) = 19$^{\mathrm{h}}$24$^{\mathrm{m}}$01.84$^{\mathrm{s}}$ and decl.~(J2000) = $+80^\circ00'34\farcs5$, with a 90\% confidence error radius of $0\farcs62$.
The BAT light curve exhibited a complex temporal structure with a duration of $T_{90}$ = 19.23 $\pm{3.82}$ s  (ref. Markwardt  et al., GCN Circular 33541),  followed by bright flaring emission detectable below 25~keV for at least 200~s after the trigger.  The same Swift GCN Circular reported the time-averaged spectrum. The best fit is by a power law with an exponential cutoff with  a photon index of 1.22 $\pm{0.29}$, and $E_{peak}$
of 99.7 $\pm{59.7}$ keV.
%The peak count rate was approximately 4375 counts~s$^{-1}$ in the 15--350~keV energy range and occurred near the trigger time. 
%The same \textit{Swift} GCN Circular reported an optical counterpart detected by \textit{Swift}/UVOT at R.A.~(J2000) = 19$^{\mathrm{h}}$24$^{\mathrm{m}}$01.84$^{\mathrm{s}}$ and decl.~(J2000) = $+80^\circ00'34\farcs5$, with a 90\% confidence error radius of $0\farcs62$.
The burst was also independently detected by the \textit{Fermi}/Gamma-ray Burst Monitor (GBM), which triggered at 14:54:47 UT, approximately 1~s before the BAT trigger. Based on $T_{90}$ duration the burst was  classified as a likely long-duration GRB \citep{Fermi2023GCN}.

In response to an automated GCN/BAT position alert, the NUTTELA-TAO began observations of the GRB at 14:55:29 UT on 2023 March 28, just 41 s after the BAT trigger, and continued until 6720 s post-trigger \citep{KomeshGCN} simultaneously in the Sloan $g\p$, $r\p$, and $i\p$\, bands  \citep{grossan2019emission,Maksut2021}. Calibration procedures involved referencing 4 bright, isolated, and unsaturated field stars from the Pan-STARRS catalog. The data are presented in Table~\ref{tab:NUTTela}.

\subsection{Later afterglow observations}
\label{sec:obs_later}

We continued afterglow observations using telescopes from the IKI GRB-FuN network \citep{ikigrbfun} and several other instruments. 
The IKI GRB-FuN is an international collaboration coordinated by the Space Research Institute of the Russian Academy of Sciences (IKI RAS) that brings together optical telescopes distributed across different geographic longitudes to provide rapid and extended follow-up observations of GRBs. The network coordinates observations following GRB alerts to improve the temporal coverage of optical afterglows, particularly during the early and intermediate phases, and facilitates the joint analysis of data obtained with participating facilities. For GRB~230328B, observations from several IKI GRB-FuN facilities extended the temporal coverage of the optical afterglow beyond the early NUTTELA-TAO observations.
The observations obtained by our collaboration and analyzed in this work are described below, and the resulting photometric measurements are listed in Table~\ref{tab:later_obs}. 

The 1.5-meter telescope AZT-33IK of the Sayan Solar Observatory (Republic of Buryatia, Mondy, Russia) observed the source starting $\sim 30$ minutes after the trigger and taking several frames in filter $R$. Observations continued for several epochs till Apr. 22, 2023 \citep{2023GCN.33530....1B}.

We observed with the Santel-400 (0.4 m) telescope of ISON-Multa observatory (Altay Republic, Russia) in the clear filter $\sim1$ hour after the trigger and detected a fading optical afterglow \citep{2023GCN.33559....1P}. We observed with the RC-36 telescope of the ISON-Kitab observatory (Qashqadaryo Region, Kitob, Uzbekistan) approximately at the same time with the same filter and obtained the upper limit \citep{2023GCN.33530....1B}.

The GROWTH-India 0.7-meter telescope \citep[GIT;][]{2022AJ....164...90K} of the Indian Astronomical Observatory (Hanle, Ladakh, India) took several $g'r'i'$-band frames starting $\sim 1.76$ hours after the burst trigger. Typical observations were 300~s long and were stacked as needed. The long duration of the observations allowed us to separate all frames into 7 epochs.

We also observed with the 0.7-meter AS-32 telescope of the Abastumani Astrophysical Observatory (AbAO; Abastumani-Kanobili, Georgia) in the $R$ filter $\sim2$ hours after the trigger \citep{2023GCN.33530....1B}.

The RC500 telescope of the Ussuriysk Astrophysical Observatory (UAFO; Primorsky Krai, Ussuriysk, Russia; observatory code C15), a division of the Institute of Applied Astronomy of the Russian Academy of Science, observed the optical source at 2.3 hours after the trigger in clear light, obtaining a stacked frame with a total exposure of 90$\times$45 s and clearly detecting the source.

The 3.6-meter Devasthal Optical Telescope \citep[DOT;][]{2023JAI....1240009P, 2024BSRSL..93..683G} of the Aryabhatta Research Institute of Observational Sciences (ARIES; Nainital, Uttarakhand, India) took several $R$-band frames starting $\sim 7$ hours after the trigger and continued observations the next night, adding one more $R$-band epoch \citep{2023GCN.33547....1R}.

The late phases of the source were observed by the 6-meter Large Altazimuth Telescope (BTA) of the Special Astrophysical Observatory (SAO; Karachai-Cherkessian Republic, Nizhnij Arkhyz, Russia). We made two observational sets in April and May 2023, using $g'$, $r'$, and $i'$ filters.

We also used the 2.6-meter Shajn telescope (ZTSh) of the Crimean Astrophysical Observatory (CrAO; Crimea, Russia) to obtain deep imaging of the site in the $R$-band with more than 2 hours of exposure at the end of April 2023.

In May 2023, we obtained two epochs of observations with the 1.5-meter AZT-22 telescope of the Maidanak Astronomical Observatory (MAO) of the Ulugh Beg Astronomical Institute (UBAI; Qashqadaryo Viloyati, Uzbekistan) in filters $R$ and $I$. 

All optical data were reduced using \textsc{iraf} \citep{iraf_2025_all}. Standard calibration procedures, including bias subtraction, dark correction, and flat-fielding, were performed with the \texttt{ccdproc} task. To improve the signal-to-noise ratio, some images were combined using \texttt{imlintran} and \texttt{imcombine}. Aperture photometry was carried out with the \textsc{apphot} package, adopting an aperture radius equal to twice the FWHM of point sources measured for each night. Photometric calibration was performed using nearby comparison stars with magnitudes from the Pan-STARRS DR1 catalog (\citealt{ChambersPanstarrs2016}; $gri$ bands) and the USNO-B1.0 catalog (\citealt{MonetUSNO2003}; $R$ band).

In addition to our measurements, we incorporated photometry reported in GCN Circulars during the first day post-trigger: OHP/T120 ($R_\mathrm{c}$) \citep{Adami2023}; GRANDMA/KNC ($R_\mathrm{c}$); CAHA/CAFOS ($R_\mathrm{c}$) \citep{Agui2023}; and the Liverpool Telescope $r'$ and $i'$ \citep{Gompertz2023}. 
These literature data were not independently reduced in this work and are explicitly identified by their corresponding references (see also in Table~\ref{tab:later_obs}).

\subsection{X-ray observations}
The \textsc{Swift}/XRT telescope observed the X-ray afterglow starting on 2023-03-28 at 14:56:38.8 UT, i.e. $\sim110.3$ s since BAT trigger \citep{Gropp2023GCN} or $\sim21$ s after the peak optical emission. The observations span a duration of $\sim$~900 ks following the initial trigger and predominately taken in the photon-counting (PC) mode. Although the XRT provides comprehensive temporal coverage across both the early and late afterglow phases, there are noticeable gaps in the time interval corresponding to the optical re-brightening phase ($10^3 -10^4$ s after the trigger).

The X-ray data in the 0.3--10~keV band used in our analysis were obtained
from the \textit{Swift} Burst Analyser service%
\footnote{\url{https://www.swift.ac.uk/burst_analyser/01162001/}
}. Instead of using the 10 keV flux density provided by the service, the final X-ray light curve was derived from the 0.3 -- 10 keV flux data and then converted to flux density using the relation: $f_X$ = $10^{23}F_X$ / $\Delta\nu_{XRT}$, where $F_X$ is the flux in erg cm$^{-2}$ s$^{-1}$, and $\Delta\nu_{XRT}$ is the XRT bandwidth in Hz. In contrast to the flux density at a single 10 keV energy obtained after the spectral modeling, the resulting average flux density in the 0.3 -- 10 keV band is not prone to systematic errors. The resulting X-ray light curve afterglow behavior is described in Sec. \ref{sec:optical_lc}.

% The XRT data processing and analysis have been described in Sec.\ref{sec:optical_lc}.

\subsection{Radio observations}
\label{sec:radio_data}

Observations at 36.8 GHz were conducted with the 22-m RT-22 radio telescope located at Simeiz \citep{Volvach2000,Volvach2012}. The receiving system employed a modulation radiometer to reduce the influence of gain-instability fluctuations in the amplification chain and atmospheric brightness-temperature variations. The measurements were performed using a differential observing mode based on sequential positioning of the source within two feed horns with orthogonal polarizations. In this scheme, the recorded signal corresponds to the difference between antenna temperatures measured in alternating telescope positions directed toward the source and adjacent reference regions of the sky. This strategy suppresses common-mode atmospheric and instrumental contributions.

Each observational session consisted of a sequence of 30–40 integrations. The final antenna temperature was determined from the average, while the uncertainty was estimated from the dispersion of the individual records. Independent registration in two polarization channels allowed us to derive total flux densities with reduced sensitivity to possible source polarization effects. Calibration of the antenna temperature scale was performed every 3–4 h using atmospheric elevation scans obtained over a wide range of zenith distances. Flux density was calibrated using reference radio sources with well-established characteristics listed in Table \ref{tab:flux_radio_calib}. The flux densities were additionally corrected for atmospheric attenuation and the elevation dependence of the effective collecting area of the telescope. The data is provided in Table \ref{tab:later_obs}. 

\begin{table}
\centering
\caption{Flux densities for calibration sources at 36.8 GHz.}
\label{tab:flux_radio_calib}
\begin{tabular}{lcccc}
\hline
Source & DR 21 & 3C 274 & NGC 7027 & 3C 286 \\ \hline
Flux density (Jy) & 18.3 & 14.3 & 5.1 & 1.56 \\ \hline
\end{tabular}
\end{table}

\subsection{Host galaxy observations}

The source host galaxy was discovered by \citet{GCNhost} $\sim3$ months after the burst. More than a year after the optical afterglow faded, we observed the host using the 6-meter Large Altazimuth Telescope (BTA) of the Special Astrophysical Observatory (SAO RAS), Russia. On 14 June 2024, 445 days after the burst, we obtained images in $g$, $r$, and $i$ filters with exposures of 1$^h$, 1$^h$, and 0.5$^h$, respectively. The photometry of the host galaxy is provided in Table~\ref{tab:hostphot}. Deep images and a seeing of 1.4$^{\prime\prime}$ allowed us not only to detect the host galaxy, but also to resolve the structure of its neighborhood. 

We continued the observations in 2024-2025 to obtain deep multicolor images of the host to estimate the redshift modeling SED using broadband photometry. In addition to deep $g$, $r$ and $i$ images from SAO/BTA, we obtained BVRI images using the 2.6-meter Shajn telescope of the Crimean Astrophysical Observatory (CrAO/ZTSh), infrared $J$ and $K$ images from the 2.5-meter telescope SAI-25 of the Caucasian Mountain Observatory (CMO/SAI-25), the $z$-band image from the 1.5-meter AZT-20 telescope of the Assy-Turgen observatory (Assy-Turgen/AZT-20) and SAI-25 telescope, and the $H$-filter image from 10-meter Gran Telescopio Canarias (GTC) equiped with EMIR infrared spectrograph. Since the source is circumpolar for both the North Caucasus and Crimea, the stacked images from the ZTSh, SAI-25, and GTC telescopes were obtained during several different nights with good weather conditions and a seeing of $\sim$ 1.4$^{\prime\prime}$-1.5$^{\prime\prime}$ to obtain better upper limits and angular resolutions. 

Near-infrared observations ($J$ and $K$ bands) were performed with the ASTRONIRCAM camera \citep{Nadjip2017} mounted on the 2.5-m telescope of the Caucasian Mountain Observatory. The primary reduction procedure is described in \cite{Tatarnikov2023}. The final images are sums of individual frames obtained with exposures of $\sim 60$ s in the $J$ band and $\sim30$ s in the $K$ band, between which the telescope was moved by 3 arcsec.

All optical host galaxy observations were reduced and calibrated in the same way as described in Sec.~\ref{sec:obs}, and IR data photometry was calibrated using the 2MASS-PSC catalog. More details are given in Sec.~\ref{sec:host_galaxy}.

\subsection{Fermi/GBM prompt $\gamma$-ray observations}

The Gamma-ray Burst Monitor (GBM) on board the Fermi Observatory consists of 12 NaI scintillation detectors, working in the (8, 850) keV range, and 2 BGO scintillation detectors, working in the (0.2, 40) MeV range \citep{meeg09}. The GBM/Fermi data used in this work were obtained from a public FTP archive\footnote{https://heasarc.gsfc.nasa.gov/FTP/fermi/}. 
GBM/Fermi triggered and located GRB~230328B at 14:54:47.43 UT on 28 March 2023 \citep{gcn33531}. We used data from NaI\_06, NaI\_07, NaI\_09, and BGO\_01 detectors in our analysis, which were the most illuminated for the source.

\begin{figure}
\centering
	\includegraphics[width=\columnwidth]{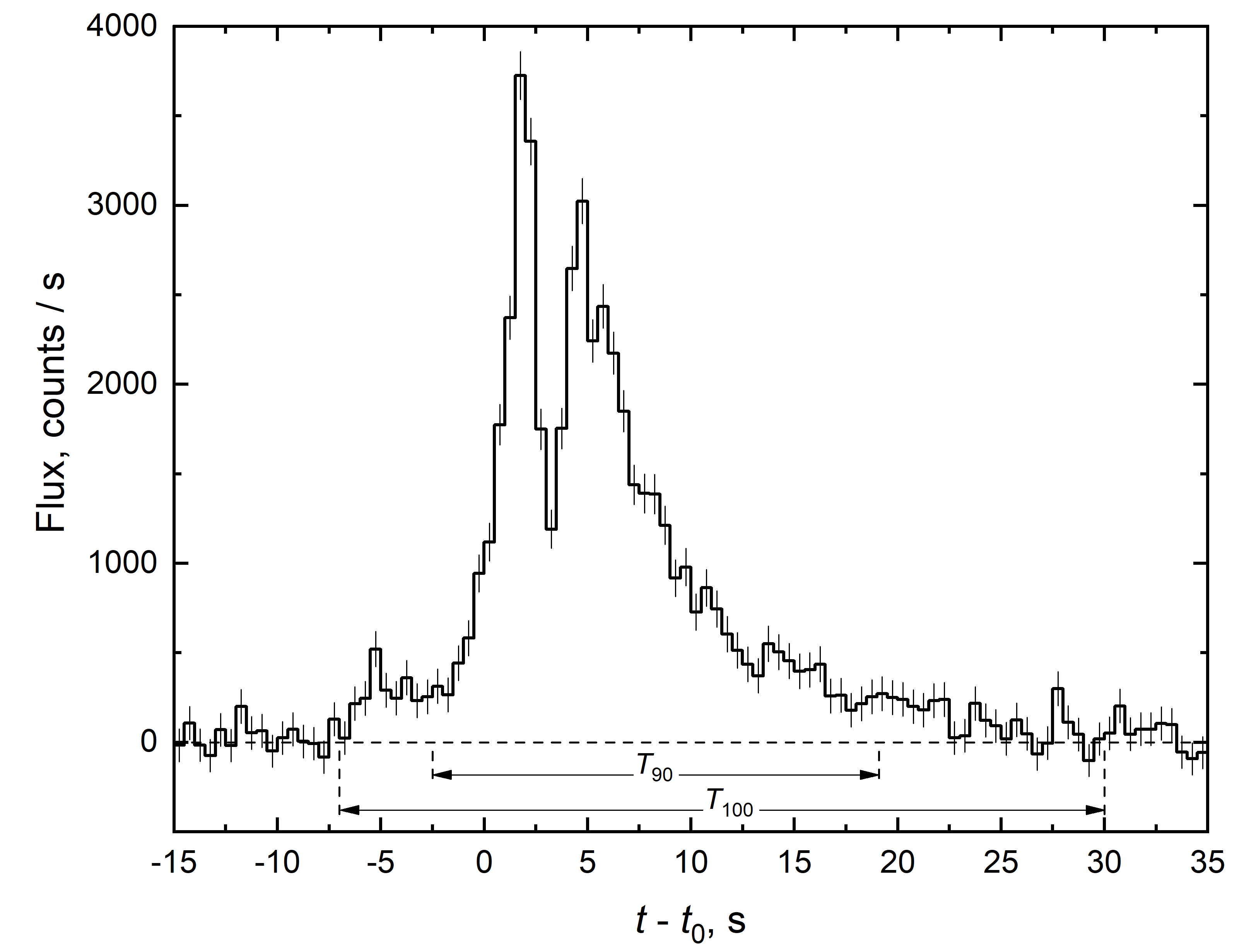}
    \caption{Background subtracted light curve of GRB~230328B based on the GBM/Fermi data from NaI\_06, NaI\_07, NaI\_09, and BGO\_01 detectors} with a time resolution of 0.5 s in the energy range of (8, 850) keV. Time intervals, covering $T_{90}$ and $ T_{100}$, are shown. 
    \label{fig:gbmlc}
\end{figure}

The background-subtracted light curve in (8, 850) keV energy range with time resolution of 0.5 s is presented in Fig.~\ref{fig:gbmlc}. We fitted the background with 3rd-order polynomial in time ranges ($-130$, $-20$)~s and (70, 470)~s jointly. The light curve could not be fit by a single exponential FRED pulse \citep{norris05}; it consists of several highly overlapped pulses. 

The burst has a total duration of $ T_{100}$ $\simeq$ 37 s, estimated using an integrated light curve with a time resolution of 0.2 s. We find that $ T_{90}$ (the time interval in which the integrated counts of the GRB increase from 5\% to 95\%) is typical for a type II (long) GRB: $T_{90}$ = 21.6 $\pm$ 0.4 s. The boundaries of the $ T_{100}$ and $T_{90}$ time intervals are shown in Fig.~\ref{fig:gbmlc}.

\begin{figure}
\centering
	\includegraphics[width=\columnwidth]{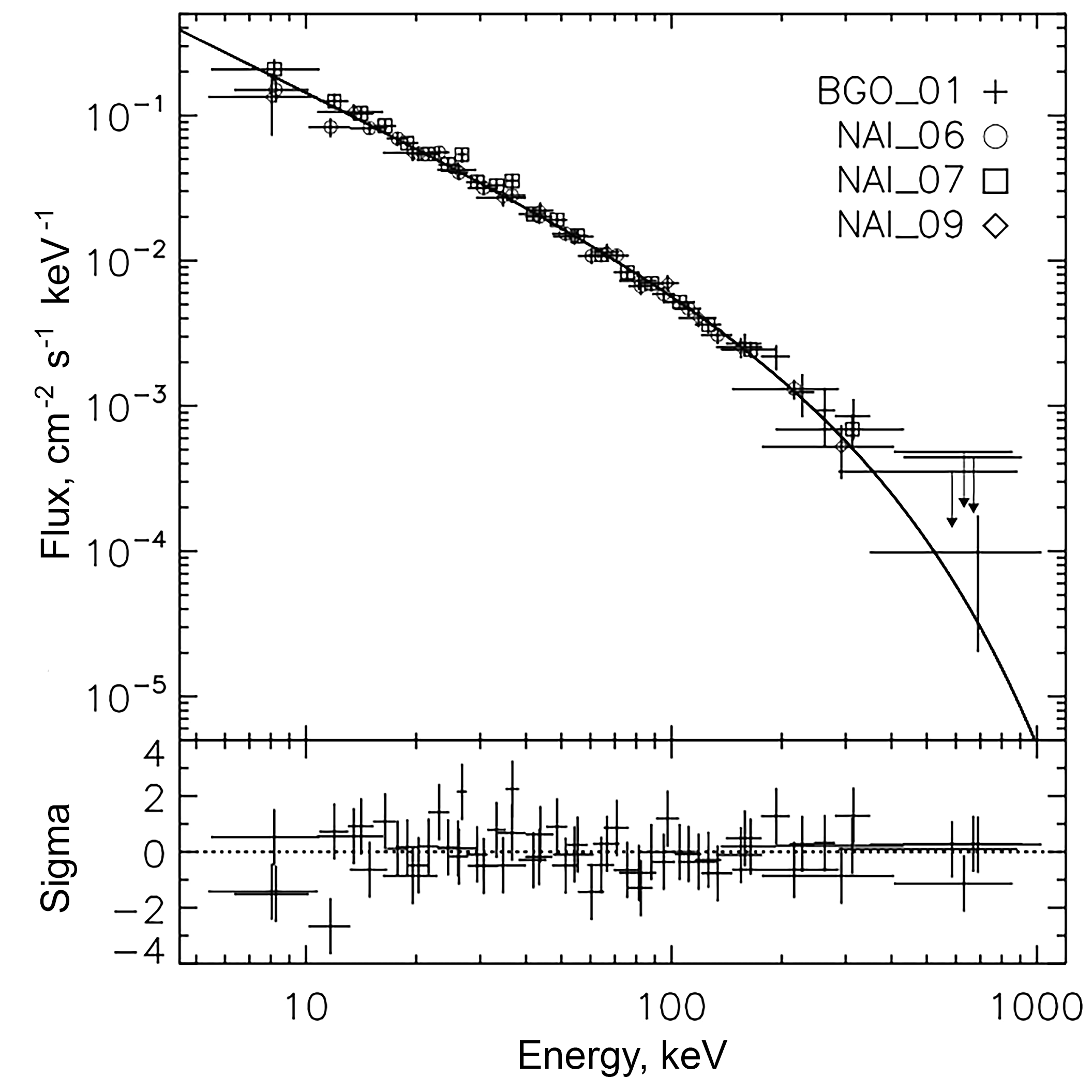}
    \caption{GRB~230328B photon spectrum obtained from the GBM/Fermi measurement in (-7, 30) s time interval with respect to the trigger time, modeled with a power-law with an exponential cutoff. The top panel shows the spectrum obtained using data of the NaI\_06, NaI\_07, NaI\_09, and BGO\_01 GBM/Fermi detectors. The bottom panel shows the deviations between the spectral model and the experimental data expressed in units of standard deviations.}
    \label{fig:gbmsp}
\end{figure}

To reconstruct the photon spectrum of the burst, we used the RMfit v4.3.2 software package, developed for analyzing data from the GBM/Fermi experiment\footnote{\url{https://fermi.gsfc.nasa.gov/ssc/data/analysis/rmfit/}}. The technique is similar to that proposed by \citet{grub14}. The spectrum was analyzed using data from the NaI\_06, NaI\_07, NaI\_09 and BGO\_01 detectors in time interval of ($-7$, 30)~s, covering the whole burst (time interval $ T_{100}$). The optimal spectral model was found to be the power law with exponential cutoff (CPL) with the following parameters: $\alpha = -1.22 \pm 0.04$, $E_\mathrm{p} = 164_{-12}^{+14}$ keV. The fluence of the burst is $S$ = (1.02 $\pm$ 0.04) $\times$ 10$^{-5}$ \,erg\,cm$^{-2}$ in (10, 1000)~keV range. The photon spectrum of GRB~230328B is shown in Fig.~\ref{fig:gbmsp}.

The long duration ($T_{90}$ = 21.6 s) and relatively soft spectrum ($E_\mathrm{p}$ = 164 keV) allow the burst to be classified as type II (long) GRB \citep[see, e.g.,][]{pool_21}.

\section{Results and discussion} 
\label{sec:res}

\subsection{Prompt gamma-ray properties}
\label{sec:prompt}
  
\begin{figure}
\centering
	\includegraphics[width=\columnwidth]{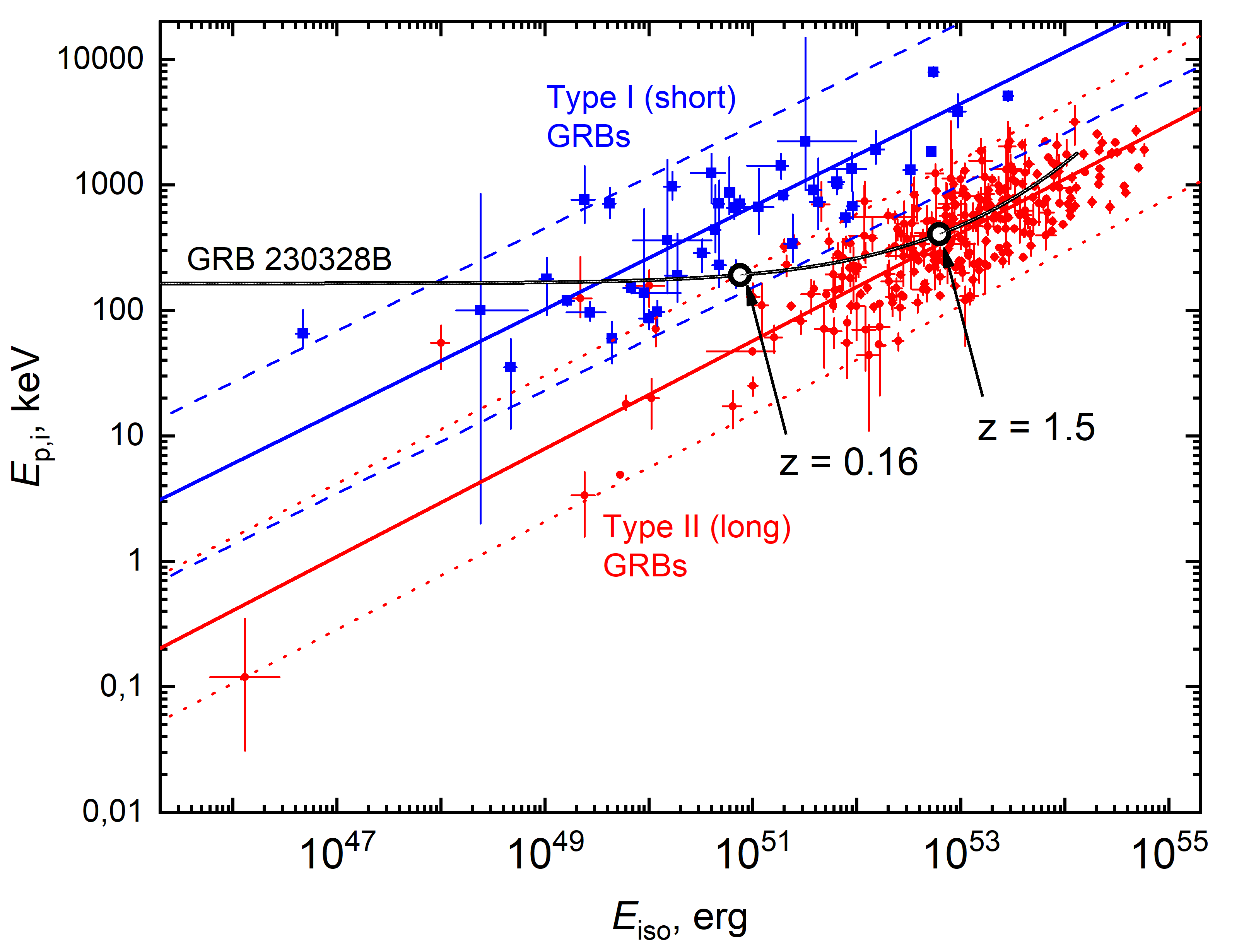}
    \caption{The $ E_\mathrm{p,i} $ -- $ E_\mathrm{iso} $ correlation for Type I (blue squares) and Type II (red circles) GRBs with the approximate results, including 2$\sigma_\mathrm{cor}$ correlation regions. The black curve traces the trajectory of GRB~230328B with redshift; the position at $z = 0.16$ and $z = 1.54$ are shown by black circles.}
    \label{fig:amati}
\end{figure}

As the redshift of GRB~230328B was not measured in direct spectroscopic observations, the isotropic energy release in the 1 keV -- 10 MeV range, $E_\mathrm{iso}$, could not be calculated. However, both the redshift and $E_\mathrm{iso}$ could be estimated using the $ E_\mathrm{p,i} $ -- $ E_\mathrm{iso} $ correlation \citep{ama02}. The correlation for the sample of 317 GRBs from \citet{min20a, min20b,min21} is shown in Fig.~\ref{fig:amati}. The trajectory of GRB~230328B in the diagram represents the dependence on the redshift. It intersects with the 2$\sigma$ correlation region of type II bursts at $z$ = 0.16, which we interpret as a lower limit of redshift. The corresponding lower limit on the total energy release in gamma rays is $ E_\mathrm{iso} $ = 7 $\times$ $10^{50}$ erg. 
The photometric redshift $z = 1.54\pm0.06$, estimated for the host galaxy of the burst (Sec.~\ref{sec:host_galaxy} and Table~\ref{tab:hostpar}), is in agreement with the $ E_\mathrm{p,i} $ -- $ E_\mathrm{iso} $ correlation for long GRBs. For $z = 1.54\pm0.06$ we obtain $ E_\mathrm{iso} $ = 6.4 $\times$ $10^{52}$ erg. 

To classify GRBs, \citet{min20a} proposed a method, which, in addition to the $ E_\mathrm{p,i} $ -- $ E_\mathrm{iso} $ correlation, uses the bimodality of the GRBs duration distribution in the rest frame $T_{90,i}$. For this, the $EH$ parameter was introduced: $EH = \frac{(E_\mathrm{p,i} / 100~\mathrm{keV})}{ (E_\mathrm{iso} / 10^{51}~\mathrm{erg})^{~0.4}}$. It characterizes the position of a GRB on the $ E_\mathrm{p,i} $ -- $ E_\mathrm{iso} $ diagram: e.g., type II GRBs are softer and brighter than type I bursts; therefore, type II bursts have a smaller value of the $EH$ parameter. 

\begin{figure}
\centering
\includegraphics[width=\columnwidth]{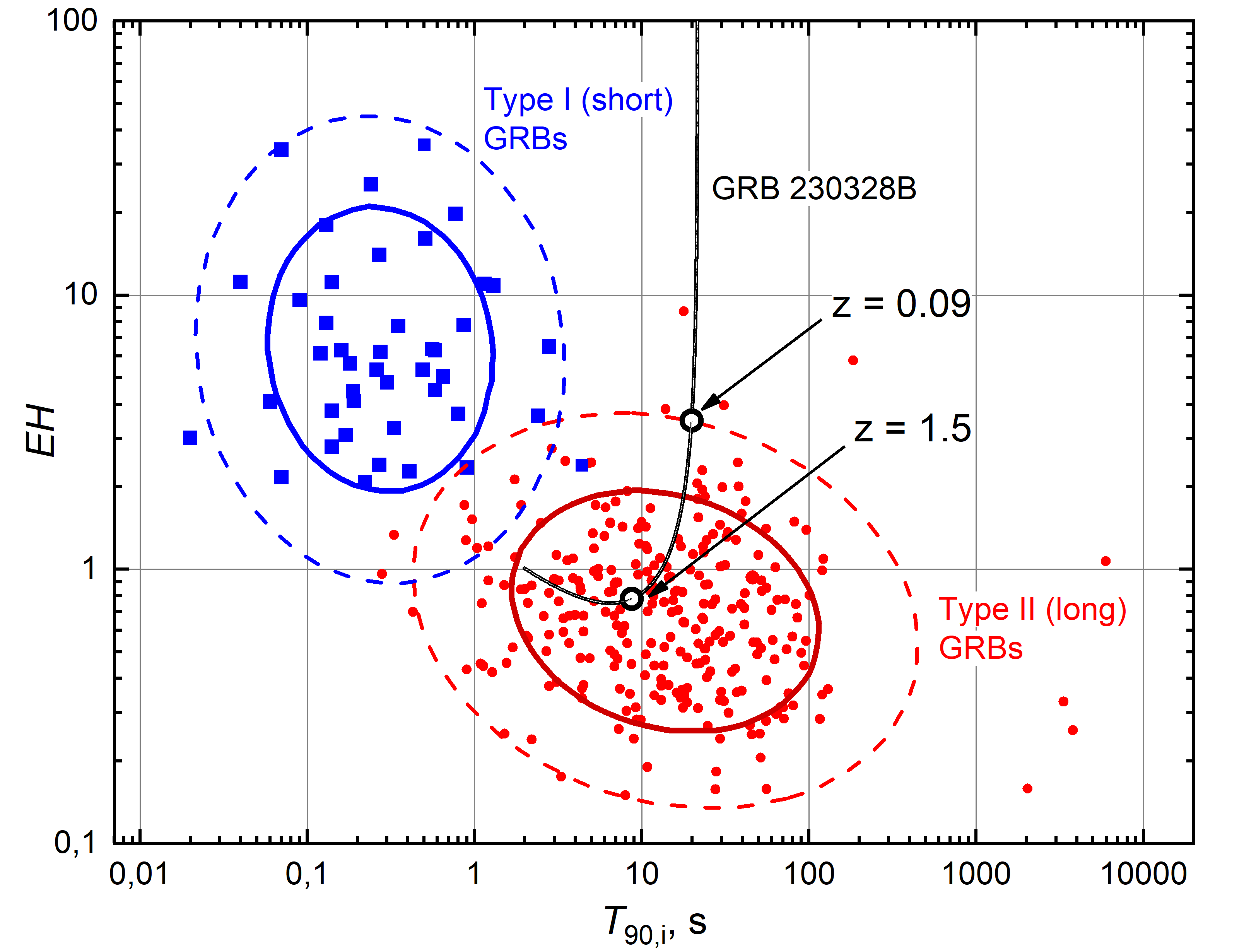}
    \caption{The $T_{\mathrm{90,i}} - EH$ diagram for Type I (blue squares) and Type II (red circles) GRBs is shown together with the corresponding cluster analysis results, with the 68\% and 95\% confidence regions indicated by bold solid and thin dashed curves, respectively. The black curve traces the trajectory of GRB~230328B with redshift, with the positions at $z = 0.09$ and $z = 1.54$ marked by black circles.
    \label{fig:ehd}}
\end{figure}

The $T_{90,i} $ -- $ EH $ diagram is shown in Fig.~\ref{fig:ehd}, constructed for the GRB sample of \cite{min20a, min20b,min21}. We plotted the trajectory of GRB~230328B in the diagram, varying its redshift. We estimated the lower limit, as we did before with the $ E_\mathrm{p,i} $ -- $ E_\mathrm{iso} $ correlation. The lower limit of redshift (intersection point of the trajectory and 2$\sigma$ cluster region of type II bursts), $z$ = 0.09 is slightly less than we obtained using the $ E_\mathrm{p,i} $ -- $ E_\mathrm{iso} $ correlation. Again, the photometrical redshift $z = 1.54\pm0.06$, measured for the host galaxy of the burst, is in good agreement with the $T_{90,i} $ -- $ EH $ diagram for long GRBs.

\begin{table*}
\centering
\caption{
Parameters obtained from MCMC fitting of the optical, radio, and X-ray light
curves with smoothly broken power-law models. The fit interval is the complete
time range used to constrain all parameters of the corresponding component.
The indices $\alpha_1$, $\alpha_3$, $\alpha_5$, and
$\alpha_{1,\mathrm{X}}$ describe the pre-peak evolution, whereas
$\alpha_2$, $\alpha_4$, $\alpha_6$, and $\alpha_{2,\mathrm{X}}$ describe the
post-peak evolution. The parameters $t_{\mathrm{p},i}$ and $\omega_i$ denote
the peak time and smoothness parameter, respectively. The optical
$\chi^2/\mathrm{d.o.f.}$ refers to the joint two-component fit.
}
\label{tab:mcmc_params}

\begin{tabular}{lllc}
\toprule
Component & Fit interval after trigger & Parameter & Value \\
\midrule

Optical onset
 & $48$--$1463$~s
 & $A_1$ & $2.27 \pm 0.51$ \\

 & & $\alpha_1$ & $1.25 \pm 0.41$ \\
 & & $\alpha_2$ & $-1.28 \pm 0.16$ \\
 & & $t_{\mathrm{p},1}$ & $89 \pm 18$~s \\
 & & $\omega_1$ & $0.67 \pm 0.20$ \\

\addlinespace
Optical rebrightening
 & $2.25\times10^3$--$1.08\times10^5$~s
 & $A_2$ & $1.00 \pm 0.03$ \\

 & & $\alpha_3$ & $1.61 \pm 0.42$ \\
 & & $\alpha_4$ & $-0.90 \pm 0.12$ \\
 & & $t_{\mathrm{p},2}$ & $3942 \pm 256$~s \\
 & & $\omega_2$ & $1.03 \pm 0.38$ \\

\addlinespace
Optical joint fit
 & $48$--$1.08\times10^5$~s
 & $\chi^2/\mathrm{d.o.f.}$ & $94/40$ \\

\midrule
Radio
 & $1.52\times10^5$--$6.81\times10^5$~s
 & $\alpha_5$ & $0.70 \pm 0.44$ \\

 & & $\alpha_6$ & $-1.22 \pm 1.03$ \\
 & & $t_{\mathrm{p},3}$ & $(4.21 \pm 1.28)\times10^5$~s \\
 & & $\omega_{\mathrm{radio}}$ & $2.59 \pm 1.52$ \\
 & & $\chi^2/\mathrm{d.o.f.}$ & $0.28/2$ \\

\midrule
X-ray
 & $3.77\times10^2$--$8.26\times10^5$~s
 & $\alpha_{1,\mathrm{X}}$ & $-0.28 \pm 0.05$ \\

 & & $\alpha_{2,\mathrm{X}}$ & $-1.43 \pm 0.03$ \\
 & & $t_{\mathrm{p},\mathrm{X}}$ & $5328 \pm 432$~s \\
 & & $\omega_{\mathrm{X}}$ & $1.13 \pm 0.22$ \\
 & & $\chi^2/\mathrm{d.o.f.}$ & $155/118$ \\

\bottomrule
\end{tabular}
\end{table*}

\subsection{Afterglow morphology}
\label{sec:optical_lc}

The optical and $\gamma$/X-ray light curve of GRB~230328B is shown in
Fig.~\ref{fig:LC1}. We can identify two main features in the optical light
curve: the onset bump with a peak at $\sim 10^2$~s and a rebrightening bump
that peaks at $\sim 4 \times 10^3$~s. We do not observe clear signatures of
\textcolor{red}{a} jet break.

To analyze the light curve, we use a two-component smoothly broken power-law
(SBPL) model, a common approach in GRB afterglow studies
\citep[e.g.,][]{Liang2007,Li2012}. The form of a single SBPL component is:
\begin{equation}\label{eq:sbpl}
f_{\rm }(t) = f_0 \left[
\left( \frac{t}{t_{\rm p}} \right)^{\alpha_1 w}
+
\left( \frac{t}{t_{\rm p}} \right)^{\alpha_2 w}
\right]^{-1/w},
\end{equation}
where $f_0$ is the normalization factor, $t_{\rm p}$ is the peak time,
$\alpha_1$ and $\alpha_2$ are the temporal indices before
and after the break, and $w$ controls the sharpness of the
transition. The full model is the sum of two such components:
\begin{equation}
F(t) = f_{\rm 1}(t) + f_{\rm 2}(t).
\end{equation}
where both $f_1$ and $f_2$ follow the form given above, but
with independent parameters. To estimate the model parameters, we
perform Markov Chain Monte Carlo (MCMC) fitting using the
Python module \texttt{emcee} \citep{Foreman-Mackey2013}. 
For the optical band, due to its extensive coverage, we
combine the $r\p$, $R$, and $R_\mathrm{c}$ bands for
fitting.

We find that $t_{\rm p}$ is largely insensitive to the
smoothness parameter $w$, while the temporal slopes depend somewhat on $w$.
Since the light curve shows smooth peaks, $w$ typically remains close to 1.
The best-fitting parameters are provided in Table~\ref{tab:mcmc_params}. The first optical bump was fitted over the 48–1463~s interval. The optical flux peak at $t_\mathrm{p} \sim 89 \pm 18$~s. The rise to the peak has a slope of $\alpha_1 = 1.25 \pm 0.43$, followed by a power-law decay with a slope of $\alpha_2 = -1.28 \pm 0.16$. The rebrightening component was fitted over $2.25\times10^3 - 1.08\times10^5$~s and rebrightening peaks at $3,942 \pm 256$~s, with rising and declining slopes of $\alpha_3 = 1.61 \pm 0.42$ and $\alpha_4 = -0.90 \pm 0.12$, respectively. 

The radio light curve was fitted over
$1.52\times10^5 - 6.81\times10^5$~s. Radio emission peaks at
$t_{\rm p,3}=(4.21\pm1.28)\times10^5$~s, with rising and declining slopes of
$\alpha_5=0.70\pm0.44$ and $\alpha_6=-1.22\pm1.03$, respectively.

The X-ray light curve was fitted over $3.77\times 10^2-8.26\times10^5$~s. The late-time declining slope of the X-ray light curve after the rebrightening is $\alpha_{2,\mathrm{X}} = -1.43 \pm 0.03$, which is somewhat steeper than the contemporaneous optical slope of $\alpha_4 = -0.90 \pm 0.12$. The shallower optical slope could be a result of the sparse optical coverage beyond a few $10^4$~s; therefore, we refrain from drawing firm
conclusions from this difference.

\subsection{Afterglow broadband spectra}
\label{sec:broadband}

\begin{table}
\caption{Results of the multiwavelength fittings to the afterglow SED obtained in the time window of $10^3-10^4$~s after the GRB trigger. The values of $E_{B-V}$, $R_V$, $A_V$, and $N_H$ for the Galaxy were fixed: $E_{B-V} = 0.069$ mag, $R_V = 3.08$, and $N_H = 0.0507 \times 10^{22}$ cm$^{-2}$. }
\centering
\small
\label{tab:SEDfit}
\begin{tabular}{lcccc}
\hline
Ext. & $E_{B-V}$ & $N_H$ & $\beta$ & $\chi^2 / d.f.$\\
law  & mag & $10^{22}$ cm$^{-2}$ & &\\
\hline
MW & $0.26 \pm 0.02$ & $0.22 \pm 0.16$ & $-0.80 \pm 0.02$ & 14.7/12\\
LMC & $0.22 \pm 0.02$ & $0.18 \pm 0.16$ & $-0.78 \pm 0.02$ & 36/12\\
SMC & $0.18 \pm 0.02$ & $0.11 \pm 0.16$ & $-0.73 \pm 0.02$ & 60/12\\
\hline
\end{tabular}
\end{table}

To construct the broadband spectra, we combined the $g\p r\p i\p$ -photometry of NUTTELA-TAO, and \textit{Swift}/XRT 0.3-10 keV spectral data in a time window $10^3 - 10^4$~s after the GRB trigger. We used HEASOFT v6.29 utilities \citep[FTOOLS,\footnote{\url{https://heasarc.gsfc.nasa.gov/ftools}}][]{blackburn1995heasoft,blackburn1999heasoft,heasarc2014heasoft}. The photometric measurements in the $g'r'i'$ filters were converted to PHA files using \texttt{flx2pha} utility.  To obtain the XRT spectra, we used the \textit{Swift} Burst Analyzer service. We grouped initial spectral channels by 25 to increase the signal-to-noise ratio of individual measurements using \texttt{grppha} tool. The spectral energy distributions (SEDs) were fitted using the absorbed power law model according to \citet{schady2010dust}, which accounts for UV/optical/IR attenuation by dust particles and photoelectric absorption of X-Ray radiation. 

We find that a broken power-law model provides a poor fit to the observed SED; therefore, the broken power law fitting results are not provided in the paper. The absorbed spectral model as implemented in \texttt{XSPEC} is given by
\begin{equation}
    \texttt{zdust} \times \texttt{zdust} \times \texttt{phabs} \times \texttt{zphabs} \times \texttt{powerlaw},
\end{equation}
where \texttt{zdust} are UV/opt/IR extinction curves \citep{pei1992dust} separately for the Galaxy and the host galaxy. \texttt{phabs} and \texttt{zphabs} are photoelectric absorption of X-rays in the Galaxy and in the host galaxy, in turn. \texttt{powerlaw} component is a power law model in the energy domain $F_E \propto E^{-\Gamma}$, where $\Gamma$ is a photon index. We fixed Galactic parameters to $E_{B-V}$ = 0.069 \citep{SF2011_Extinction} and $N_H = 0.0507 \times 10^{22}$ cm$^{-2}$, calculated using built-in absorption maps in \texttt{XSPEC}. The redshift for the Galactic component \texttt{zdust} was fixed at $z$ = 0, while for the host \texttt{zdust} the redshift was set at $z = 1.54$, as estimated in Sec.~\ref{sec:host_galaxy}. The ISM element abundance \texttt{wilm} \citep{wilmXrayAbund2000} was chosen instead of the Solar one. We tested three different host galaxy extinction curves: the Milky Way (MW), the Large Magellanic Cloud (LMC), and the Small Magellanic Cloud (SMC). The fit was performed with the Levenberg-Marquardt method. The results of our multi-wavelength SED fitting are summarized in Table~\ref{tab:SEDfit}. 

\begin{figure}
  \centering
  \includegraphics[width=\columnwidth]{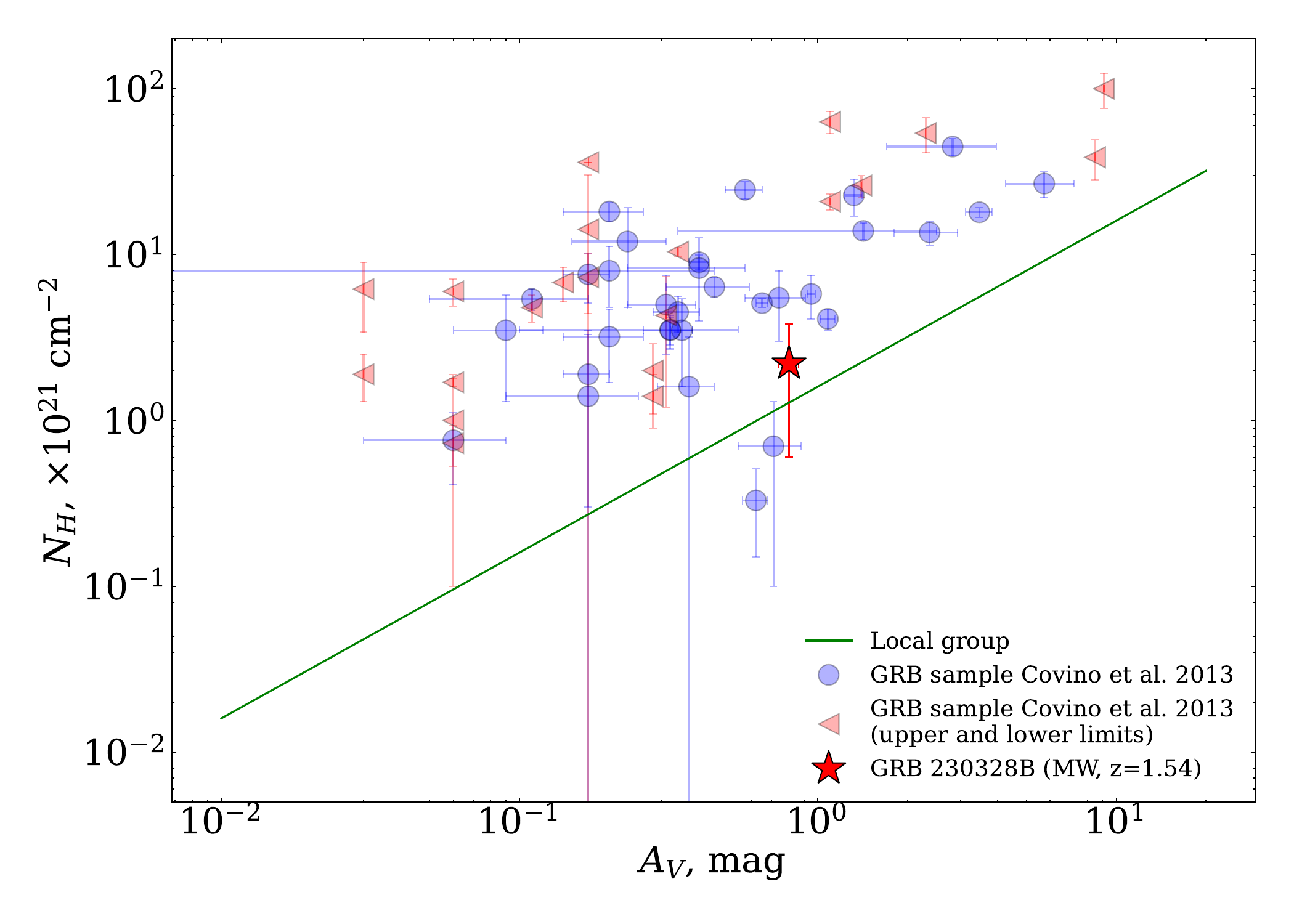}
  \caption{The dust-to-gas ratio of GRB~230328B host (red star; obtained for the MW extinction law) in comparison with the sample of the GRB hosts from \citet{Covino2013Dust}. Blue dots are values and red triangles are upper or lower limits (colors markers were chosen as in the original paper). The green line indicates the dust-to-gas ratio for the Local group of galaxies for which $N_H/A_V \sim 1.6 \times 10^{21}$ cm$^{-2}$ mag$^{-1}$.}
\label{fig:grb_230328b_host_av_nh_comparison}
\end{figure}

Among the tested extinction laws, the MW and LMC curves provide better fits to the data, with lower reduced chi-square values compared to the SMC law. This suggests that the host galaxy dust properties in the line-of-sight may better resemble MW or LMC dust distributions rather than the steeper UV extinction characteristic of the SMC.
The MW extinction $A_V =0.80 \pm 0.06$ mag, the reddening $E_{B-V} = 0.26 \pm 0.02$, and the hydrogen column density $N_H = (0.22 \pm 0.16) \times 10^{22}$~cm$^{-2}$ are in good agreement with the values obtained from host galaxy analysis (Sec. \ref{sec:host_galaxy}). Additionally, we compared the estimated absorption parameters with those listed in the sample of GRB hosts presented in \citet{Covino2013Dust}.  Fig.~\ref{fig:grb_230328b_host_av_nh_comparison} illustrates the position of the GRB~230328B host on the $N_H$ -- $A_V$ correlation for the best fit parameters. 

This indicates that the host of GRB~230328B is highly absorbed. Such properties are consistent with a spiral galaxy, as proposed in Sec.~\ref{sec:host_galaxy}, with characteristics similar to the LMC or the MW. At $z \sim 1.54$, the inferred absorption may place the burst in an active star-forming region. Alternatively, the high absorption could arise if the GRB is located on the far side of the host disk, resulting in a large optical depth along the line of sight.

\subsection{Afterglow modeling}
\label{sec:afterglow}

\begin{table}
\centering
\caption{Parameters for GRB~230328B obtained from the afterglow MCMC fit using an energy injection model, quoted as posterior medians with $1\sigma$ uncertainties. $E_{\mathrm{k,iso}}$ and $\Gamma_0$ denote the isotropic-equivalent kinetic energy and initial jet Lorentz factor. $p$ is the electron power-law index, while $\varepsilon_{\mathrm{e}}$ and $\varepsilon_{\mathrm{B}}$ are the electron and magnetic-field energy fractions. The circumburst medium is described by density $n_0$ and profile index $k$, and energy injection by $f_{\mathrm{inj}}$ and $\ell$. Finally, $R_0$ and $\Delta R/R_0$ specify the initial radius and relative shell thickness.}
\label{tab:best_fit_params}
\begin{tabular}{l c @{\qquad} l c}
\toprule
Parameter & Value & Parameter & Value \\
\midrule
$\log_{10} E_\mathrm{k,iso}$ & $53.92^{+0.20}_{-0.18}$ & $k$ & $0.39^{+0.16}_{-0.17}$ \\
$\log_{10} \Gamma_0$          & $3.50^{+0.28}_{-0.34}$  & $A_V$ & $0.48^{+0.05}_{-0.05}$ \\
$\log_{10} n_0$               & $-1.02^{+0.78}_{-0.84}$ & $f_\mathrm{inj}$ & $3.29^{+0.61}_{-0.56}$ \\
$p$                           & $2.39^{+0.03}_{-0.03}$  & $\ell$ & $0.02^{+0.02}_{-0.01}$ \\
$\log_{10} \varepsilon_e$     & $-1.91^{+0.18}_{-0.19}$ & $\log_{10} R_0$ & $17.70^{+0.27}_{-0.25}$ \\
$\log_{10} \varepsilon_B$     & $-3.70^{+0.61}_{-0.57}$ & $\Delta R/R_0$ & $1.73^{+0.31}_{-0.30}$ \\
\bottomrule
\end{tabular}
\end{table}

\begin{figure}
    \centering
    \includegraphics[width=1.0\linewidth]{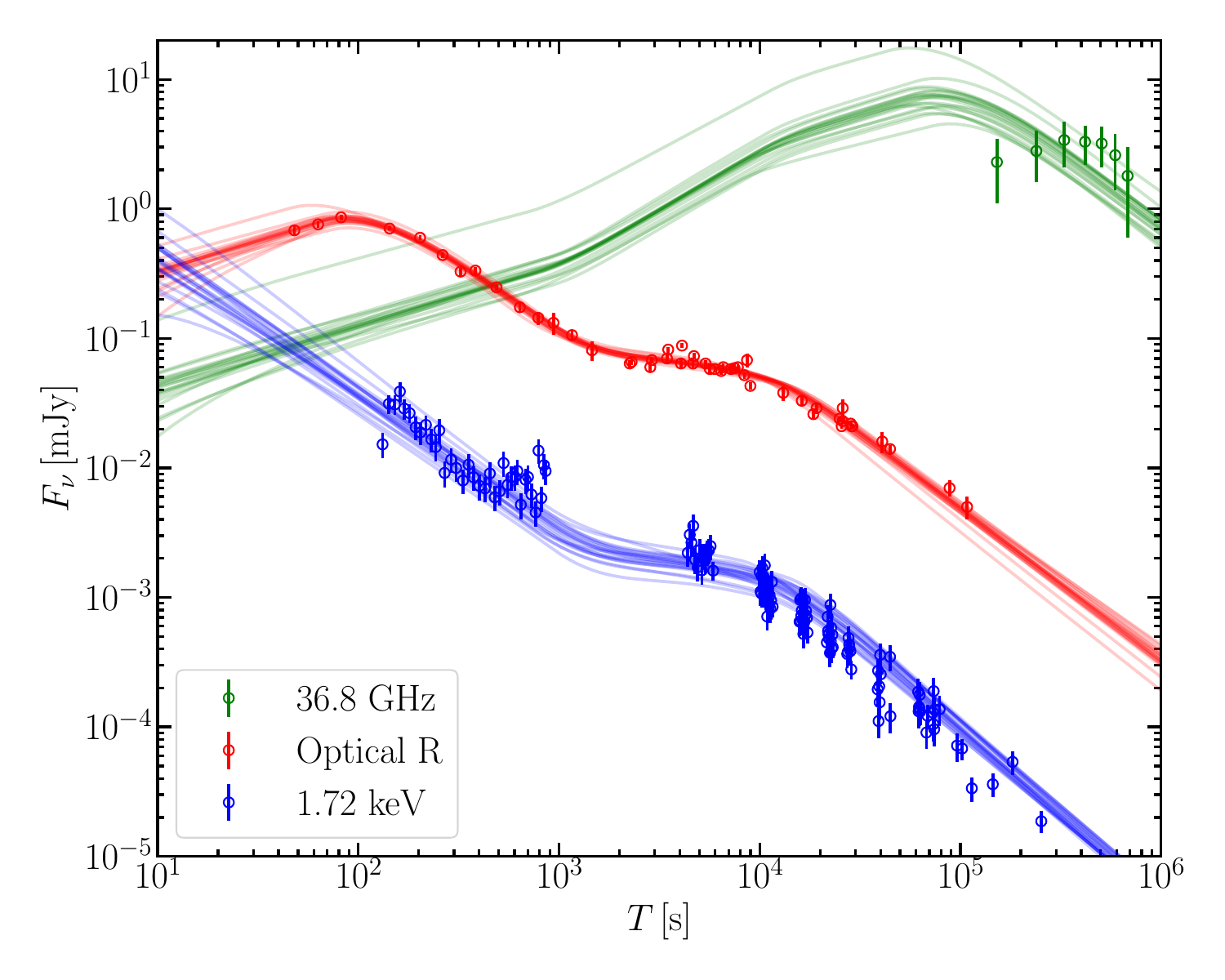}
    \caption{The light curves obtained from MCMC analysis together with observed data points in the radio, optical R, and X-ray bands, shown with green, red, and blue dots. See Table~\ref{tab:best_fit_params} for the values of the best fit parameters.}
    \label{fig:LC_MCMC}
\end{figure}

\begin{figure}
    \centering
    \includegraphics[width=1.0\linewidth]{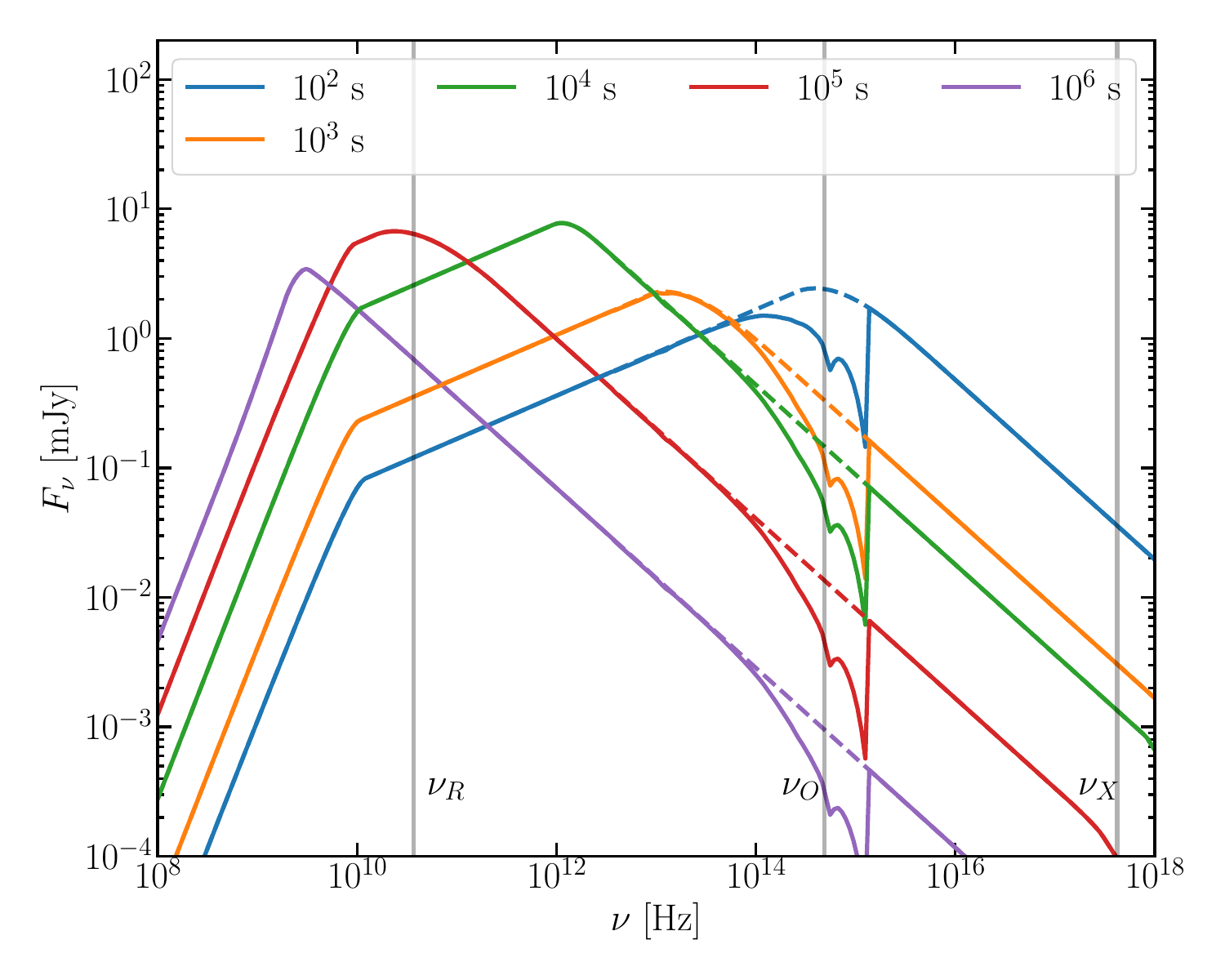}
    \caption{Afterglow spectra corresponding to best-fit parameters obtained from MCMC analysis at various epochs. The solid lines include the extinction of the host, while dashed lines do not. The vertical gray lines indicate the frequencies at which observations were obtained.}
    \label{fig:spectrum_MCMC}
\end{figure}

We perform an MCMC fit to the afterglow data using the numerical code of \citet{Gill-Granot-18,Gill23GRB}. To model the optical extinction, we use the public Python library \texttt{dust\_extinction} \citep{Gordon-24} with the MW extinction law (Sec.~\ref{sec:broadband}).

As noted above, the GRB~230328B afterglow exhibits a rebrightening bump. Such bumps are commonly observed in GRBs \citep[e.g.,][]{liang2013comprehensive, Wang15How, Mazaeva2018, AnguloValdez26Evidence, Akl26Multi, Fu26fast}, and several mechanisms have been proposed to explain them \citep{Barkov2011MNRAS, Zhang24BOAT}. These include two-component jets, where the onset bump originates from the jet core and the rebrightening from a wider jet wing \citep[e.g.,][]{Peng05Two, Mukazhanov26Systematic}; late energy injection \citep{Sari00Impulsive, Zhang02GRB, Laskar15Energy, AnguloValdez26Evidence, Fu26fast}; long-lived reverse shocks \citep{Uhm12Dynamics}; structured jets \citep[e.g.,][]{Berger03common, Jia12Physical}; and off-axis viewing geometries \citep[e.g.,][]{Abdikamalov25Reverse}. 

Here we explore the energy injection model recently used by \citet{AnguloValdez26Evidence}. This model considers a spherical blast wave with initial isotropic-equivalent kinetic energy $E_{\rm k,iso}$ and coasting bulk Lorentz factor $\Gamma_0$. Additional energy, $E_{\rm inj}=f_{\rm inj}E_{\rm k,iso}$, is injected into the expanding blast wave starting at radius $R_0$ and ending at $(1+\Delta R/R_0)R_0$. This energy is added gradually as a power law,
$$
\frac{dE}{dR} = \frac{(1 + \ell)}{\Delta R} E_{\mathrm{inj}} \left( \frac{R - R_0}{\Delta R} \right)^\ell \propto R^\ell\,,
$$
where $\ell$ is a parameter that characterizes the energy injection rate.
Thus, the total blast wave energy grows as $E(R)\propto R^{1+\ell}$. 

The MCMC fit result is given in Table~\ref{tab:best_fit_params}. The light curve is shown in Fig.~\ref{fig:LC_MCMC} and the corner plot is in Fig.~\ref{fig:cp_MCMC}. The model describes the X-ray and optical emission well, but it is less successful in reproducing the radio signal. The peak in the modeled radio light curve is caused by the passage of the injection (characteristic) frequency $\nu_m$, and occurs slightly earlier than the observed peak. Since the afterglow model is agnostic to how energy is injected into the blast wave, it is possible that the shock microphysical parameters, which are kept fixed, might change during this process, delaying the peak to later times. Alternatively, the radio emission might be coming from a separate emission component, e.g.\ the reverse shock that may arise due to the refreshing of the shock by slower-moving shells behind the outer decelerating shell. 

Our MCMC fit converges on a solution where the initial blast wave is highly relativistic ($\Gamma_0 \approx 3160$) and energetic ($E_{\mathrm{k,iso}} \approx 8.3 \times 10^{53}$~erg). It propagates through an external medium with a shallow density profile ($n \propto R^{-k}$) where $k \approx 0.39$ and a density normalization $n_0 \approx 0.10$ at a radius of $10^{18}$\,cm. Such intermediate density profiles ($0<k<2$), departing from canonical homogeneous ISM ($k=0$) or pure wind ($k=2$) environments, are frequently reported in afterglow modeling \citep[e.g.,][]{Tian22Constraining}, but the physics behind it is not entirely clear. The dynamical parameters of the blast wave and the properties of the external medium dictate the radius at which the blast wave decelerates after sweeping up mass $M_{\rm sw}=M_0/\Gamma_0=E_{\rm k,iso}/\Gamma_0^2c^2$ in the thin-shell approximation \citep{sari95hydrodynamics}, where $M_0$ is the baryon load of the ejecta. The corresponding observer-frame time for the deceleration is $T_{\rm dec} = (1+z)[(3-k)E_{\rm k,iso}/(2^{5-k}\pi Ac^{5-k}\Gamma_0^{2(4-k)})]^{1/(3-k)}\simeq0.13$\,s when using the best-fit parameters from our model. This timescale is much shorter than the duration of the prompt emission that has a $T_{90}=21.6$\,s. The thin-shell approximation demands that $T_{\rm dec}>T_{90}$, which is achieved when $\Gamma_0\lesssim500$, which is more typical, while keeping all other model parameters fixed. Therefore, the large $\Gamma_0$ obtained in the fit should be treated as an upper limit rather than a true value, where the latter remains uncertain due to the missing peak of the X-ray light curve that is generally used to constrain $T_{\rm dec}$. For $T>T_{\rm dec}$ the dynamical evolution of the blast wave becomes self-similar and therefore independent of $\Gamma_0$. 

When comparing the prompt isotropic-equivalent $\gamma$-ray energy ($E_{\gamma,\mathrm{iso}} = 6.4 \times 10^{52}$~erg) to the initial kinetic energy, we estimate a prompt radiative efficiency of $\eta_\gamma\equiv E_{\gamma,\rm iso}/(E_{\gamma,\rm iso}+E_{\rm k,iso}) \approx 7\%$. This value is consistent with expectations for internal energy dissipation mechanisms in GRB jets \citep{Fan06Gamma,Beniamini+16}. The true kinetic energy of the bipolar jets depends on the beaming correction that yields $E_{\rm k}=(1-\cos\theta_j)E_{\rm k,iso}\simeq\theta_j^2E_{\rm k,iso}/2\simeq4.1\times10^{51}\theta_{j,-1}^2$\,erg, which is standard when assuming a typical jet opening angle of $\theta_j=0.1\theta_{j,-1}$\,rad. When the $1/\Gamma$ angular size of the beaming cone of radiation around the observer's line-of-sight, for which we assume $\theta_{\rm obs}=0$, grows beyond $\theta_j$ as the blast wave slows down, the afterglow light curve shows an achromatic steepening, commonly referred to as a \textit{jet break}, due to missing emission from outside of the jet core. The jet break time for the assumed jet structure and viewing geometry is given by $T_j=(\Gamma_0\theta_j)^{2(4-k)/(3-k)}T_{\rm dec}\simeq12.5$\,days, however, the late-time afterglow observations are insufficient to confirm the presence of a jet break.

The microphysics parameters describe a weakly magnetized forward shock ($\varepsilon_B \approx 2.0 \times 10^{-4}$, $\varepsilon_e \approx 0.012$), which fall within the typical ranges inferred from broad afterglow sample studies \citep[e.g.,][]{Yi13early, Wang15How} The inferred electron power-law index ($p=2.39 \pm 0.03$) is tightly constrained and aligns well with the canonical values \citep{liang2013comprehensive, Li18Large}. The host extinction parameter is found to be $A_V = 0.48^{+0.05}_{-0.05}$.

The rebrightening is driven by the energy injection process, characterized by an injection factor $f_{\mathrm{inj}} \approx 3.29$ and temporal index $\ell \approx 0.02$. Energy injection commences at $R_0 \approx 5.0 \times 10^{17}$~cm and continues over a fractional radial width of $\Delta R/R_0 \approx 1.73$. The amount of injected energy is significant, but not unprecedented and has been required at a similar or higher levels to explain such rebrightenings in other GRBs \citep[e.g.][]{Laskar+18,Genevieve+24,deWet+24,Genevieve+25,AnguloValdez26Evidence}. As a result, the total kinetic energy budget of the blast wave now becomes $E_{\rm k}=(1+f_{\rm inj})E_{\rm k,0}\simeq1.8\times10^{52}\theta_{j,-1}^2$\,erg, where $E_{\rm k,0}$ is the kinetic energy of the initial blast wave before energy injection. The total energy budget is at the high end of the distribution but remains consistent with earlier estimates. For example, \citet{Bloom-Frail-Kulkarni-2003} find a mean beaming corrected prompt $\gamma$-ray energy for a sample of 29 GRBs to be $\langle E_\gamma\rangle\simeq1.33\times10^{51}$\,erg, and when assuming a typical $\gamma$-ray efficiency of $\eta_\gamma=0.1$, the mean kinetic energy is $\langle E_{\rm k}\rangle\simeq1.2\times10^{52}$\,erg. The energy of the blast wave gradually grows over radius after injection and correspondingly over observer-frame time, such that $E_{\rm k}(T)\propto T^{(1+\ell)/[3-(\ell+k)]}$ (see, e.g., \citealt{Gill+2026} for scalings). We can compare our parameterized energy-injection model to a particular scenario, e.g., when the inner ejecta that energizes the outer (initial) ejecta has a radial velocity stratification, i.e. $M(\Gamma)\propto\Gamma^{-s}$, so that more mass and kinetic energy resides in the slower moving ejecta \citep{Sari00Impulsive}. In this scenario, the blast wave energy grows as $E_{\rm k}(T)\propto T^{-(3-k)(1-s)/(7+s-2k)}$ that yields $s = (4-k+\ell)/(2-k-\ell) = 2.3$ when comparing it with our model.

Spectra from this model are shown in Fig~\ref{fig:spectrum_MCMC} at various epochs. The radio emission remains above the self-absorption frequency ($\nu_a$) at all times. The optical emission lies in the slow-cooling regime, $\nu_{\mathrm{m}} < \nu_O < \nu_{\mathrm{c}}$, at all times. The cooling frequency $\nu_{\mathrm{c}}$ starts entering the X-ray band at $\simeq10^5$\,s. The spectrum is consistent with the finding from the  SED fit (Sec.~\ref{sec:broadband}) that it can be fitted with a single optical–X-ray power law over $10^3$–$10^4$ s. 

To further support this, the observed optical spectral slope ($\beta$; Fig.~\ref{fig:LC1}, upper panel) from simultaneous $g'r'i'$ NUTTELA-TAO photometry yields an inverse-variance weighted mean of $\bar{\beta}=-2.00\pm0.10$ across three intervals ($\beta_{41\text{--}93\mathrm{s}}=-2.03\pm0.18$, $\beta_{113\text{--}1613\mathrm{s}}=-2.10\pm0.15$, $\beta_{2004\text{--}6900\mathrm{s}}=-1.78\pm0.21$). A constant-slope fit ($\chi^2=1.57$, 2~d.o.f., $p=0.46$) indicates no significant color evolution over $\simeq10^2$--$7\times10^3$~s. Later Liverpool $r'i'$ observations at $4.46\times10^4$~s are modestly bluer ($\beta=-1.29\pm0.32$, a $\sim2\sigma$ difference), but are limited to two bands. We therefore adopt a nearly constant optical continuum, supporting our model finding that the intrinsic afterglow optical emission remains in the slow-cooling regime for most of its evolution.

While our data are consistent with the energy injection interpretation, several caveats apply. Most importantly, we fit a 12-parameter model, and the high dimensionality makes it difficult to fully exclude local minima. Furthermore, the early X-ray data exhibit flaring, and the lack of contemporaneous radio coverage limits the constraints on parameters shaping the early afterglow. Finally, while energy injection naturally explains the rebrightening, we cannot completely rule out alternative physical scenarios, such as a structured two-component jet \citep{Mukazhanov26Systematic}.

\subsection{Host galaxy}
\label{sec:host_galaxy}

\begin{table*}
\caption{
Photometry of the host galaxy. In the first three columns filter name, effective average wavelength, and effective FWHM are listed. $A_\mathrm{MW}$ shows the value of the Galactic extinction subtracted from the magnitude of the host galaxy for every filter, based on the extinction maps from \citet{SF2011_Extinction}.} $T_\mathrm{exp}$ is the exposure time. UL gives the $3\sigma$ upper limit sensitivity in magnitudes for images of the given exposure time.
\label{tab:hostphot}
\centering
\begin{tabular}{@{}crrlrccl@{}}
\hline
Filter & $\lambda_\mathrm{eff}$, \r{A} & $\Delta\lambda$, \r{A} & $A_\mathrm{MW}$ & $T_\mathrm{exp}$ [s] & mag (err) & UL & Obs./Tel.\\
\hline
$B$ & 4370  & 820  &  0.247 & 27120 & $-$          & 24.0 & CrAO/ZTSh \\
$g$ & 4920  & 710  &  0.225 & 3600  & 24.33 (0.06) & 26.0 & SAO/BTA \\
$V$ & 5420  & 890  &  0.183 & 13560 & $-$          & 24.0 & CrAO/ZTSh \\
$r$ & 6540  & 900  &  0.156 & 3600  & 23.89 (0.07) & 26.1 & SAO/BTA \\
$R$ & 6600  & 1570 &  0.148 & 14640 & 24.03 (0.08) & 25.4 & CrAO/ZTSh \\
$i$ & 7540  & 1300 &  0.116 & 1800  & 23.48 (0.09) & 25.1 & SAO/BTA \\
$I$ & 7600  & 1210 &  0.103 & 5520  & 23.71 (0.15) & 24.0 & CrAO/ZTSh \\
$z$ & 8960  & 990  &  0.086 & 2700  & 23.31 (0.16) & 23.9 & CMO/SAI-25 \\
$J$ & 12410 & 2150 &  0.048 & 7200  & 22.46 (0.17) & 23.2 & CMO/SAI-25 \\
$H$ & 16290 & 2810 &  0.031 & 16200 & $-$          & 21.8 & EMIR/GTC \\
$K$ & 21660 & 2780 &  0.021 & 7200  & 22.31 (0.27) & 22.4 & CMO/SAI-25 \\
\hline
\end{tabular}
\end{table*}

\begin{figure}
\centering
\includegraphics[width=\linewidth]{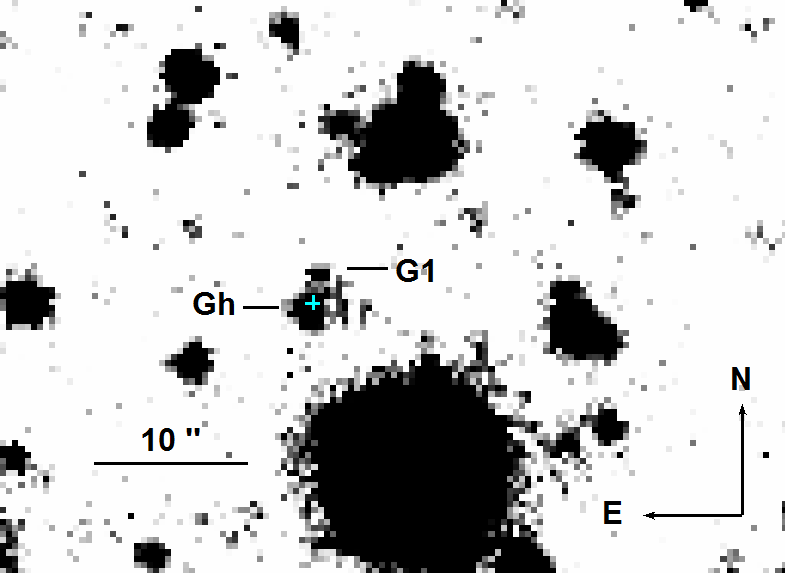}
    \caption{The localization region of the GRB~230328B observed by the 6-m BTA telescope of SAO RAS 445 days after the burst trigger, $gri$ combined image. The cross indicates the position of the GRB optical afterglow detected by \textit{Swift}/UVOT at R.A.~(J2000) = 19$^{\mathrm{h}}$24$^{\mathrm{m}}$01.84$^{\mathrm{s}}$, Dec.~(J2000) = $+80^\circ00'34\farcs5$ with an uncertainty of 0\farcs62 \citep{Gropp2023GCN}. The host galaxy is marked as Gh. A possible interacting neighbor galaxy is marked as G1.}
    \label{fig:host_map}
\end{figure}

Figure~\ref{fig:host_map} shows the stacked image of the $gri$ filters taken 445 days after the burst trigger. The localization of the source is marked by a cross. The nearest visual environment of the source position is complex, revealing the conglomerate of possibly several interacting galaxies. To avoid contamination, we studied only the brightest part of the structure, marked in Fig.~\ref{fig:host_map} as Gh, assuming that this is the host galaxy of GRB~230328B. Assuming the coordinates of the optical afterglow from \citet{Gropp2023GCN} with the uncertainty of 0\farcs62 and taking the galaxy brightness from CrAO/ZTSh filter $R = 24.03 \pm 0.08$ we estimate the probability of chance association between the host galaxy and the optical transient using Eq.(1)-(3) from \citep{Bloom2002} as $P_\mathrm{ch}$ = 0.52\%. Thus we can conclude that the source Gh is the host galaxy of the GRB~230328B.

The host galaxy Gh is clearly detected with the neighboring structure in all bands except $B$, $V$, and $H$. The photometry of the host was made using \texttt{APPHOT} task of the NOAO’s \textsc{IRAF} software package\footnote{\texttt{https://iraf-community.github.io/x11iraf.html}}. To minimize the contribution of neighboring sources and to collect equivalent flux for every filter, we chose a fixed circular aperture with the diameter of 3.5 arcsec centered on the brightness center of the Gh source from the stacked BTA $g'r'i'$ image, i.e., at R.A.~(J2000) = 19$^{\mathrm{h}}$24$^{\mathrm{m}}$02.03$^{\mathrm{s}}$, Dec.~(J2000) = $+80^\circ00'33\farcs9$ (with statistical error of 0\farcs5). The offset of the afterglow position \citep{Gropp2023GCN} from the host galaxy center lies within the statistical errors of the measurements of $0\farcs8$. Hence, we can place only the upper limit of the offset of 7 kpc. The photometry of the host galaxy is listed in Table~\ref{tab:hostphot}. All magnitudes were corrected for the Galactic reddening using maps of extinction from \citet{SF2011_Extinction}. 

\begin{figure}
\centering
\includegraphics[width=\linewidth]{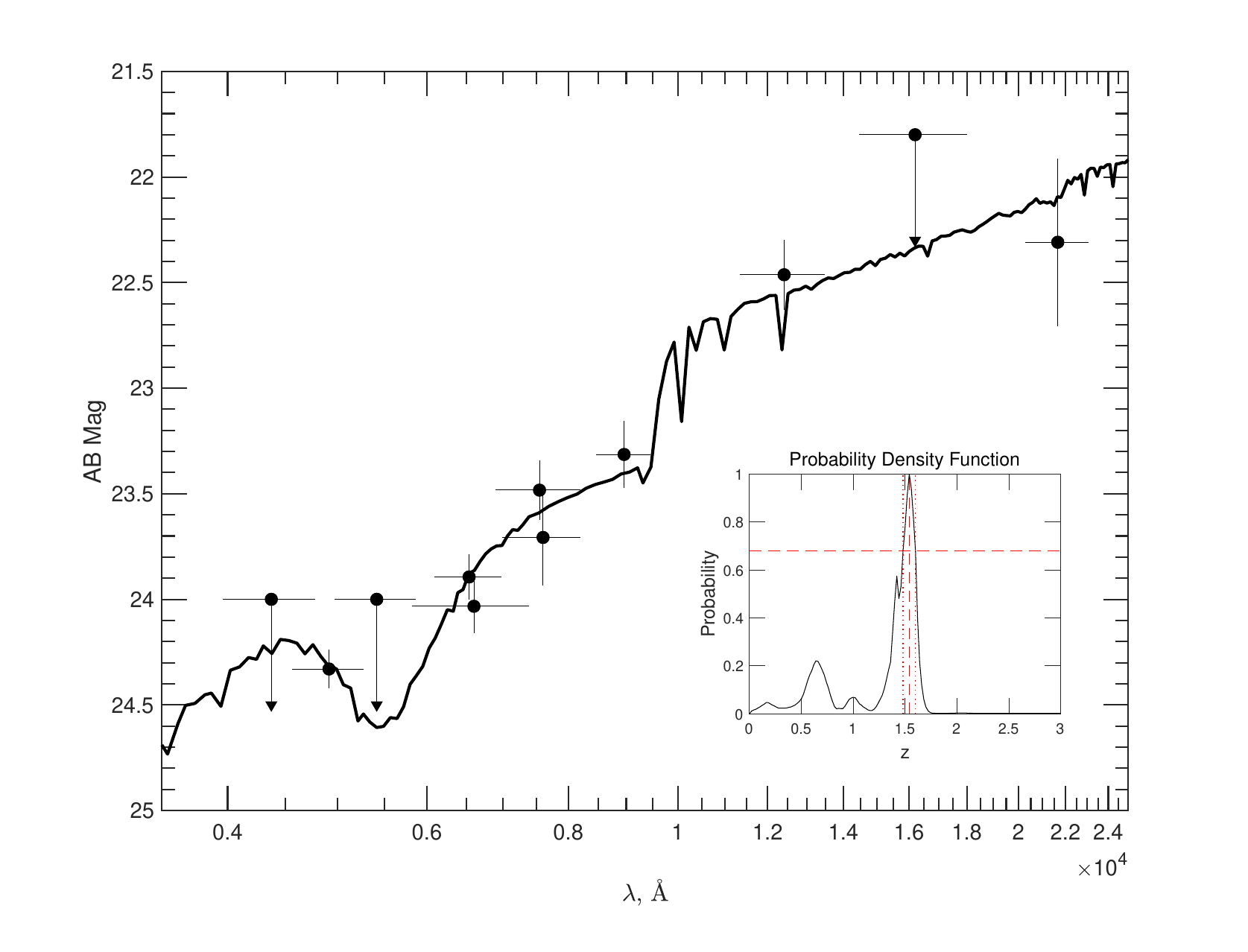}
\caption{The host galaxy spectral energy distribution modeled with the broadband photometry for the redshift estimation. Circles depict the extragalactic magnitudes of the host in the filters $BgVrRIizJHK$. Y-axis error-bars are $1\sigma$ photometric errors, X-error-bars are filter wights magnitudes determined as $\mathrm{FWHM}/2$ of the bandwidth of the filter. Arrows mark upper limits. Solid line is the best modeled host SED from the \textsc{PEGASE2} library for the S0-type galaxy at the redshift $z=1.54\pm0.06$ with $E_{B-V}\sim0.2$. The inset displays the probability density function (solid line) of the model of the redshift, the horizontal dashed line depicts 1$\sigma$ (68\%) level, the vertical dashed line marks the maximum probability redshift for the model, and the two dotted lines mark 1$\sigma$ boundaries of the best-modeled redshift.}
\label{host:sed}
\end{figure}

We used the \textsc{Le~Phare} software package~\citep{lephar1,lephar2} to model the broadband photometry of the host with a synthetic SED. We used the \textsc{PEGASE2} library of population synthesis models \citep{pegase} to obtain the best-fitted SED and the main physical parameters of the galaxy: its type, redshift, average bulk extinction, age, mass, and star formation rate. According to the results, the host galaxy is a relatively young S0-type galaxy with a redshift of $z=1.54\pm0.06$ with a size similar to the Milky Way: assuming the angular size of the photometry aperture of 3.5$^{\prime\prime}$ as the host angular size, we estimate its physical dimension as $\sim 28$ kpc\footnote{For distances calculations, we used the NED Cosmology Calculator by \citet{Wright2006} with $H_{0} = 69.6$ km~s$^{-1}$~Mpc$^{-1}$, $\Omega_{M} = 0.286$, and $\Omega_{\Lambda} = 0.714$.}. All host parameters are listed in Table~\ref{tab:hostpar}. Fig.~\ref{host:sed} shows the best-fitted host SED along with the corresponding photometry of the galaxy and the probability density function for the best model ($\chi^2/DOF=4.4/8$).

\begin{table}
\caption{Physical parameters of the host galaxy Gh.}
\label{tab:hostpar}
\centering
\begin{tabularx}{\columnwidth}{@{}lrcl@{}}
\hline
Redshift & $z$ & = & $1.54 \pm 0.06$ \\
Type & S0 &  &  \\
Diameter & $d$ & $\sim $ & $28$ kpc \\
Abs. $R$ magnitude$^a$ & $M_R$ & = & -22.0 \\
Bulk extinction$^b$ & $E_{B-V}$ & $\sim$ & 0.2 \\
Age & $t$ & = & $7.1 \times 10^8$ yr \\
Mass & $M$ & = & $1.6 \times 10^{10}$ M$_{\bigodot}$ \\
Star form. rate & $\mathrm{SFR}$ & = & 35 M$_{\bigodot}/$yr \\
\hline
\end{tabularx}
\footnotesize{}
\begin{tablenotes}
\item [a] $^a$ $R$-filter in the observer frame.
\item [b] $^b$ assuming the Milky Way extinction law from \citet{seaton}.
\end{tablenotes}
\end{table}

Deep SAO RAS/BTA imaging of the host galaxy region (Fig.~\ref{fig:host_map}) reveals a much fainter component G1 (the estimated flux is $\sim 25 \%$ of that of Gh, i.e., 1.5 magnitude fainter than Gh based on the BTA $gri$-sum aperture instrumental photometry of the two objects) located northwest of Gh. If indeed this source G1 is a neighboring galaxy at an angular distance of about 2.5 arcsec between the centers of the galaxies, then the system could be an interacting pair of galaxies at a distance of $\sim 20$ kpc at $z=1.54$. Some long GRB hosts show ongoing star formation, and interacting systems are not unprecedented: several GRBs have been found in galaxies exhibiting clear signs of interaction or merging \citep[see, e.g.,][]{volnova051008, jelinek210312B}.

\begin{figure}
\centering
\includegraphics[width=\linewidth]{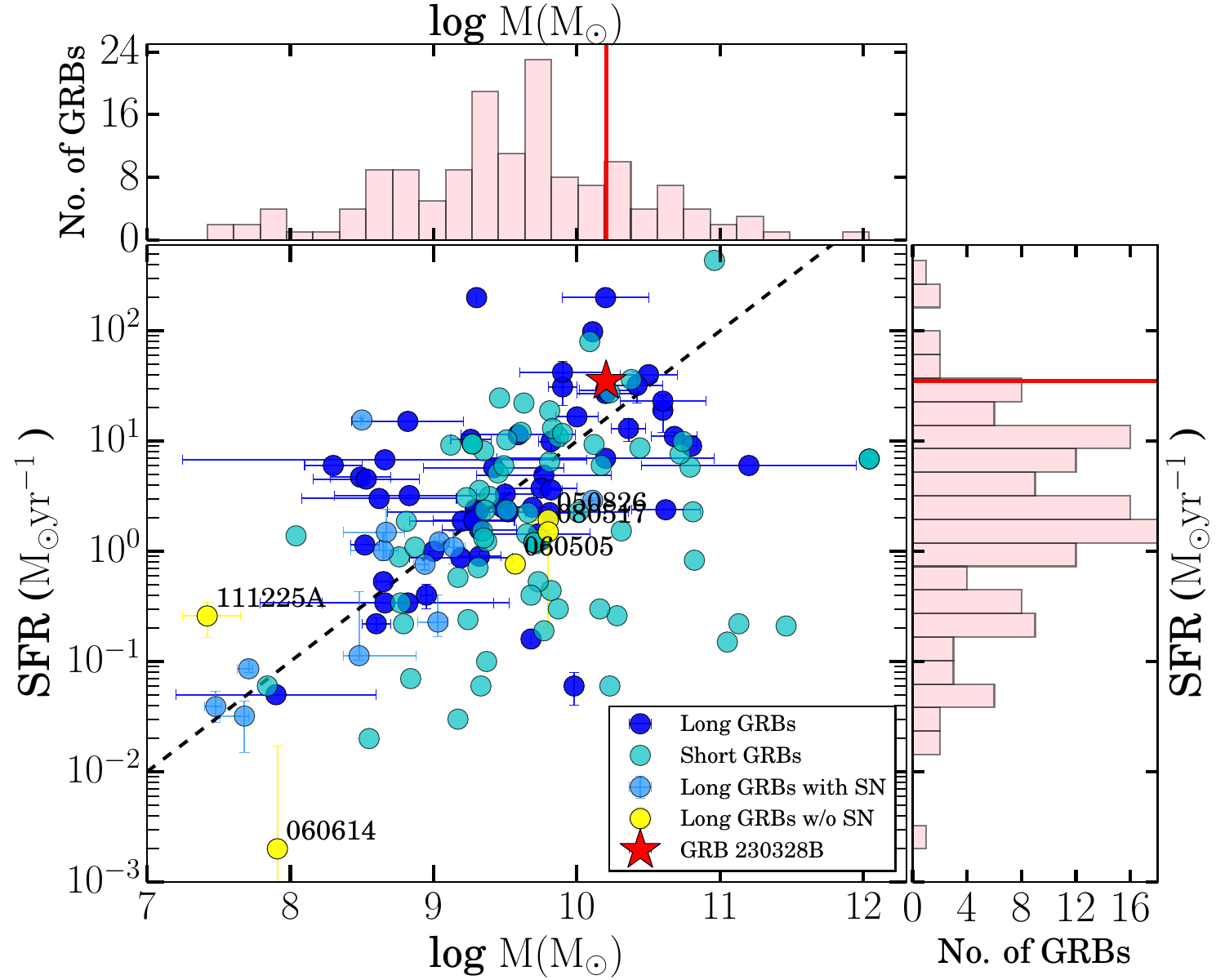}
\caption{Stellar mass versus star formation rate (SFR) for the host galaxy of GRB 230328B (red star) compared to a compiled sample of GRB host galaxies, including long GRBs (blue circles), short GRBs (cyan circles), long GRBs with associated supernovae (light blue circles with cross-hatching), and long GRBs without detected supernovae (yellow circles; GRBs 050826, 060505, 060614, 080517, and 111225A). The dashed line represents the star-forming main sequence. Marginal histograms show the distributions of stellar mass (top) and SFR (right) for the full comparison sample; the red vertical and horizontal lines mark the values derived for GRB 230328B. The host galaxy of GRB 230328B occupies the high-mass, very high-SFR end of the distribution.}
\label{fig:mass_sfr}
\end{figure}

Figure~\ref{fig:mass_sfr} shows the position of the host galaxy of GRB~230328B in the stellar mass--star formation rate plane, compared to a compiled sample of long GRBs, short GRBs, long GRBs with associated supernovae, and long GRBs without detected supernovae \citep[e.g.,][]{2006AIPC..836..540S, Savaglio2009, 2021MNRAS.503.3931T, 2022ApJ...940...57N, Gupta22Gamma}. 
With $\log(M_{\star}/\mathrm{M_{\odot}}) \approx 10.2$ and $\mathrm{SFR} \approx 35~\mathrm{M_{\odot}~yr^{-1}}$, the host of GRB~230328B  occupies the high-mass, actively star-forming end of the long GRB host distribution, lying above the star-forming main sequence  \citep[dashed line;][]{Elbaz11}. These properties are characteristic of  massive, collapsar-type GRB hosts \citep{Kruhler2015}.

\subsection{Searching for supernova}
\label{sec:SN}

It is well known that some long-duration GRBs are accompanied by Type Ic supernovae~\citep[e.g.,][]{Finneran25GRBSN, Aimuratov23GRBSN, HuSN}. However, supernova searches are generally effective only at low redshift, as at higher redshifts the sensitivity of current instruments is often insufficient to detect a supernova against the host galaxy background \citep{Gupta2025arXiv}.

The redshift of GRB~230328B was not determined during the afterglow phase. Nevertheless, we conducted a search for a supernova signature, as each additional detection provides valuable constraints on the population of GRB-associated supernovae. We monitored the GRB localization region from 3 to 48 days after the trigger (Table~\ref{tab:later_obs}).

No statistically significant excess in the source brightness (i.e., host galaxy plus a potential supernova) was detected in any of the filters ($g^\prime, R, r^\prime, i^\prime$) over this period. The photometry remained consistent across epochs, allowing us to place upper limits on any supernova contribution. Using the deepest observations, we constrained the maximum brightness a supernova could have had without being detected. We utilized BTA/SAO RAS observations obtained at 18 and 48 days post-trigger. The positive 3$\sigma$ photometric uncertainties at these epochs provide upper limits on the total flux from the host galaxy and any supernova component. From these data, we derive limiting magnitudes of $i^\prime = 23.2$ at 18.3 days and $i^\prime = 23.1$ at 48.4 days. At the redshift of $z = 1.54$, these values correspond to absolute magnitudes of $-22.0$ in the $u^\prime$ filter at 7.3 and 19.4 days in the GRB rest frame. (These magnitudes are not corrected for the host galaxy extinction.) 

A natural question is whether GRB~230328B could belong to the rare class of long-duration GRBs lacking an associated supernova. The confirmed SN-less long-duration GRBs, most notably GRB~060614 \citep{Gehrels06, Fynbo06} --- form a clearly distinct sub-population in Figure~\ref{fig:mass_sfr}, residing exclusively in low-mass ($\log(M_{\star}/\mathrm{M_{\odot}}) \lesssim 10$), low-SFR dwarf galaxies, separated from the host of GRB~230328B by a few orders of magnitude in both stellar mass and SFR. The host of GRB~230328B bears no resemblance to this class in any of its global properties, strongly disfavouring an SN-less long GRB classification based on host galaxy evidence alone.

We note that at $z = 1.54$, our late-time photometric limits are too shallow to place meaningful constraints on a potential supernova component associated with GRB~230328B. The derived limiting absolute magnitude lies well above the peak luminosities of known GRB-associated supernovae \citep{Belkin_2020AstL...46..783B,Belkin_2024AstL}, therefore, these limits cannot be used to classify GRB~230328B as a genuinely SN-less event. The argument against such a classification thus rests entirely on the host galaxy properties described above. The high stellar mass and elevated SFR are instead consistent with a collapsar progenitor embedded in a massive, dusty galaxy where a contemporaneous supernova could be substantially attenuated below the detection threshold at the observed redshift.

\section{Conclusion}
\label{sec:conc}

We present multi-wavelength observations of GRB~230328B, spanning from 41~s post-trigger to deep host-galaxy imaging over a year later. This enables us to probe the physical mechanisms of the burst, characterize its environment, and place it within the broader population of long-duration GRBs. Our main findings are summarized below.

The prompt emission observed by Fermi/GBM shows a multi-pulsed light curve with $T_{90} = 21.6 \pm 0.4$~s. Spectral fitting with a cutoff power law yields peak energy $E_\mathrm{p} = 164_{-12}^{+14}$~keV. The long duration and soft spectrum classify it as a typical long GRB. Using the photometric redshift $z \approx 1.54$, we infer $E_\mathrm{iso} \approx 6.4 \times 10^{52}$~erg. In the rest frame, the burst is consistent with standard long-GRB correlations, including the Amati relation and the $T_{90,i}$–$EH$ diagram (Sec.~\ref{sec:prompt}).

Combined optical and X-ray modeling reveals significant line-of-sight absorption, with $A_V = 0.80 \pm 0.06$~mag. Broadband fitting indicates that the host extinction is best described by Milky Way or LMC-like curves rather than SMC-like dust. The high extinction suggests the burst occurred in or behind a dense star-forming region within the host galaxy, implying substantial intervening material along the line of sight (Sec.~\ref{sec:broadband}).

The optical and X-ray afterglow exhibits a complex morphology characterized by an onset bump and an achromatic optical rebrightening at about 4000 s. MCMC modeling reveals a highly relativistic and energetic initial blast wave with initial energy of $E_{\mathrm{k,iso}} \approx 8.3 \times 10^{53}$~erg. This corresponds to a radiative efficiency of $\approx 7\%$, which aligns with standard expectations for energy dissipation mechanisms in GRB jets. As the jet propagates through a low-density, weakly magnetized external medium, the late-time rebrightening can be explained by gradual energy injection from trailing shells. While this model reproduces the key afterglow features, the complexity of the multi-parameter fit, early X-ray flaring, and a lack of contemporaneous early radio coverage introduce some uncertainties. Therefore, alternative physical scenarios, such as a structured two-component jet, cannot be ruled out (Sec~\ref{sec:afterglow}).

Based on observations taken 445 days after the burst, the host is identified as a relatively young S0-type galaxy at $z \approx 1.54$. Spanning approximately 30 kpc, it is comparable in size to the Milky Way and resides in a complex environment, likely part of interacting galaxies. The burst afterglow was localized close (within 7 kpc) to this host galaxy center. This central location, combined with the heavy dust absorption observed, strongly suggests that the event occurred deep within one of these highly obscured, active star-forming regions (Sec.~\ref{sec:host_galaxy}).

We monitored the burst region from 3 to 48 days post-trigger to search for an associated supernova, finding no variability or excess emission. At the estimated redshift $z \approx 1.54$, our limits are sensitive only to supernovae brighter than typical GRB-associated events. Thus, any typical accompanying supernova would be too faint to detect against the host galaxy background (Sec.~\ref{sec:SN}).

Overall, GRB~230328B provides a valuable look at a massive stellar collapse within a dense, dusty galactic environment at $z \approx 1.54$. The complex behavior of its afterglow and the strong dust absorption observed here underline the importance of future multi-band observations. Observing GRBs rapidly across multiple wavelengths will be essential for separating the physics of the burst itself from the influence of its surrounding environment.

\section*{Acknowledgements}

We dedicate this work to the memory of Bruce Grossan, whose contributions were invaluable to this project. 

We thank Bing Zhang for useful discussions. This research was funded by the Science Committee of the Ministry of Science and Higher Education of the Republic of Kazakhstan (Grant No. AP26103591). TK acknowledges support from Chinese National Foreign Expert Individual Project (Project No. 110000352520258002) and the CAS-ANSO Fellowship (Grant No. CAS-ANSO-FA-2024-06). EA is supported by the Nazarbayev University Faculty Development Competitive Research Grant Program (no. 040225FD4713). MK is supported by the Science Committee of the Ministry of Science and Higher Education of the Republic of Kazakhstan (Grant No. BR21881880). RG was sponsored by the National Aeronautics and Space Administration (NASA) through a contract with ORAU. PB's work was funded by a NASA grant 80NSSC24K0770, a grant (no. 2020747) from the United States-Israel Binational Science Foundation (BSF), Jerusalem, Israel and by a grant (no. 1649/23) from the Israel Science Foundation. 
The NUTTELA-TAO Team expresses gratitude to the personnel of the Assy-Turgen Astrophysical Observatory and the Fesenkov Astrophysical Institute, Almaty, Kazakhstan, for their assistance.
This work made use of data supplied by the UK \textit{Swift} Science Data Centre at the University of Leicester.

AP, NP, AV, AM thank the RTTAC committee for providing time for observing on the BTA-6m and CMO-2.5m telescopes. EK is grateful to the Ministry of Science and Higher Education of the Russian Federation for financial support of this work; the AZT-33IK telescope is part of the Core Shared Research Facility "Angara" of ISTP SB RAS. OAB and AP thank the Ministry of Higher Education, Science and innovation of the Republic of Uzbekistan for its support, grant no. IL-5421101855.
Observations of GRB 230328B were completed with the 22-meter radio telescope RT-22 of CrAO working 
at the frequency of 36.8 GHz using modulation radiometers (A.E. Volvach, L.N. Volvach, M.G. Larionov Characterizing 
Close Binary Supermassive Black Holes in Blazars 3C 273, 3C 454.3, S 0528+134, and AO 0235+164: Insights from 
Multi-Frequency Radio Monitoring and Gravitational Wave Prospects // The Astrophysical Journal, 2025. Vol.992, P.60. DOI: 10.3847/1538-4357/adfee2).
The GROWTH India Telescope (GIT) is a 70-cm telescope with a 0.7-degree field of view, set up by the Indian Institute of Astrophysics (IIA) and the Indian Institute of Technology Bombay (IITB) with funding from  Indo-US Science and Technology Forum and the Science and Engineering Research Board, Department of Science and Technology, Government of India. It is located at the Indian Astronomical Observatory (Hanle), operated by IIA. We acknowledge funding by the IITB alumni batch of 1994, which partially supports operations of the telescope. 
GCA acknowledges support from the Indian National Science Academy (INSA) under the INSA Senior Scientist programme.

%%%%%%%%%%%%%%%%%%%%%%%%%%%%%%%%%%%%%%%%%%%%%%%%%%
\section*{Data Availability}

The data underlying this project is available from authors upon request.

\section*{Affiliation}
$^{1}$Energetic Cosmos Laboratory, Nazarbayev University, Astana 010000, Kazakhstan\\
$^{2}$State Key Laboratory of Radio Astronomy and Technology, Xinjiang Astronomical Observatory, CAS, 150 Science 1Street, Urumqi, Xinjiang, 830011, China\\
$^{3}$Institute of Experimental and Theoretical Physics, Al-Farabi Kazakh National University, Almaty 050040, Kazakhstan\\
$^{4}$Space Research Institute, Russian Academy of Sciences, Moscow 117997, Russia\\
$^{5}$Faculty of Physics, Higher School of Economics, Moscow 101000, Russia\\
$^{6}$Instituto de Radioastronom\'ia y Astrof\'isica, Universidad Nacional Aut\'onoma de M\'exico, Antigua Carretera a P\'atzcuaro $\#$ 8701, \\ Ex-Hda. San Jos\'e de la Huerta, Morelia, Michoac\'an, C.P. 58089, M\'exico\\
$^{7}$Astrophysics Research Center of the Open university (ARCO), The Open University of Israel, PO Box 808, Ra’anana 4353701, Israel\\
$^{8}$Space Sciences Laboratory, University of California, Berkeley, CA 94720, USA\\
$^{9}$National Astronomical Observatories,Chinese Academy of Sciences, Beijing 100101, China\\
$^{10}$University of Chinese Academy of Sciences, Beijing 100080, China\\
$^{11}$School of Physics \& Astronomy, Monash University, Clayton VIC 3800, Australia\\
$^{12}$Fesenkov Astrophysical Institute, Almaty, 050020, Kazakhstan\\
$^{13}$Special Astrophysical Observatory, Russian Academy of Sciences, Nizhny Arkhyz 369167, Russia\\
$^{14}$Department of Physics, Nazarbayev University, 53 Kabanbay Batyr ave, Astana 010000, Kazakhstan\\
$^{15}$Institute of Solar-Terrestrial Physics, Russian Academy of Sciences (Siberian Branch), Irkutsk 664033, Russia\\
$^{16}$Lomonosov Moscow State University, Moscow 119991, Russia\\
$^{17}$Crimean Astrophysical Observatory, Russian Academy of Sciences,  Nauchny 298409, Crimea\\
$^{18}$Ulugh Beg Astronomical Institute, Uzbekistan Academy of Sciences, Tashkent 100052, Uzbekistan\\
$^{19}$Samarkand State University, 15, University boulevard, Samarkand 140104, Uzbekistan\\
$^{20}$Evgeni Kharadze Georgian National Astrophysical Observatory, Abastumani-Kanobili 0301, Georgia\\
$^{21}$Samtskhe-Javakheti State University, Akhaltsikhe 0080, Georgia\\
$^{22}$Keldysh Institute of Applied Mathematics of the Russian Academy of Sciences, Moscow 125047, Russia\\
$^{23}$Institute of Applied Astronomy of Russian Academy of Science, 10 Kutuzova Emb., Saint-Petersburg 191187, Russia\\
$^{24}$Aryabhatta Research Institute of Observational Sciences (ARIES), Manora Peak, Nainital, Uttarakhand 263001, India\\
$^{25}$Astrophysics Science Division, NASA Goddard Space Flight Center, Mail Code 661, Greenbelt, MD 20771, USA\\
$^{26}$NASA Postdoctoral Program Fellow\\
$^{27}$Department of Physics, Indian Institute of Technology Bombay, Powai, 400 076, India\\
$^{28}$Indian Institute of Astrophysics, 2nd Block 100 Feet Rd, Koramangala Bangalore, 560 034, India\\
$^{29}$Instituto de Astrof\'{\i}sica de Andaluc\'{\i}a -- CSIC, Glorieta de la Astronom\'{\i}a s/n, E-18008 Granada, Spain\\
$^{30}$Universit\'e de la Côte d'Azur, Nice, France\\
$^{31}$IJCLab, Univ Paris-Saclay, CNRS/IN2P3, Orsay, France\\
$^{32}$Department of Natural Sciences, The Open University of Israel, PO Box 808, Ra’anana 4353701, Israel\\
$^{33}$Department of Physics, The George Washington University, 725 21st Street NW, Washington, DC 20052, USA



\bibliographystyle{mnras}
\bibliography{GRBs} % if your bibtex file is called example.bib

%\begin{thebibliography}{99}
%\end{thebibliography}

%%%%%%%%%%%%%%%%%%%%%%%%%%%%%%%%%%%%%%%%%%%%%%%%%%

%%%%%%%%%%%%%%%%% APPENDICES %%%%%%%%%%%%%%%%%%%%%

\appendix

\renewcommand{\thefigure}{A\arabic{figure}}
\setcounter{figure}{0}                      % Restart from 0
\renewcommand{\thetable}{A\arabic{table}}  % Set table label format
\setcounter{table}{0}                      % Reset table numbering

\begin{table*}
\centering
\caption{NUTTELA-TAO observations. $t_\mathrm{mid}$ is the midpoint time of each co-add, and $t_0$ is the BAT trigger time, which occurred at 14:54:48 UT on 28 March 2023 \citep{Gropp2023GCN}. $T_\mathrm{exp}$ denotes the exposure time of each image. The Milky Way extinction is subtracted.\label{tab:NUTTela}}
\setlength{\tabcolsep}{4pt}
\begin{adjustbox}{max width=\textwidth}
\begin{tabular}{c c cc cc cc}
\hline
$t_\mathrm{mid}-t_0$ [s] & $T_\mathrm{exp}$ [s] &
\multicolumn{2}{c}{$g^\prime$ filter} &
\multicolumn{2}{c}{$r^\prime$ filter} &
\multicolumn{2}{c}{$i^\prime$ filter} \\
\cline{3-4}\cline{5-6}\cline{7-8}
 & & mag (err) & Flux (err) [mJy] & mag (err) & Flux (err) [mJy] & mag (err) & Flux (err) [mJy] \\
\hline
48    & 15   & 17.25 (0.13) & 0.457 (0.056) & 16.81 (0.08) & 0.683 (0.047) & 16.08 (0.09) & 1.338 (0.107) \\
63    & 15   & 16.90 (0.10) & 0.633 (0.056) & 16.70 (0.07) & 0.760 (0.046) & 15.96 (0.08) & 1.494 (0.112) \\
82    & 22   & 17.05 (0.12) & 0.550 (0.060) & 16.56 (0.05) & 0.861 (0.042) & 16.12 (0.08) & 1.298 (0.091) \\
143   & 60   & 17.25 (0.09) & 0.455 (0.037) & 16.78 (0.04) & 0.706 (0.024) & 16.26 (0.06) & 1.139 (0.066) \\
203   & 60   & 17.51 (0.13) & 0.360 (0.042) & 16.96 (0.04) & 0.599 (0.024) & 16.62 (0.07) & 0.815 (0.053) \\
263   & 60   & 17.73 (0.11) & 0.294 (0.029) & 17.29 (0.06) & 0.440 (0.022) & 16.97 (0.08) & 0.593 (0.043) \\
323   & 60   & 18.09 (0.14) & 0.211 (0.027) & 17.61 (0.08) & 0.328 (0.023) & 17.08 (0.08) & 0.536 (0.041) \\
383   & 60   & 18.19 (0.15) & 0.192 (0.027) & 17.59 (0.08) & 0.334 (0.023) & 17.38 (0.10) & 0.406 (0.038) \\
488   & 150  & 18.34 (0.10) & 0.167 (0.015) & 17.91 (0.06) & 0.248 (0.015) & 17.44 (0.08) & 0.382 (0.027) \\
638   & 150  & 18.96 (0.17) & 0.094 (0.014) & 18.30 (0.09) & 0.174 (0.014) & 17.60 (0.08) & 0.332 (0.025) \\
788   & 150  & 18.92 (0.15) & 0.098 (0.014) & 18.50 (0.13) & 0.144 (0.017) & 17.84 (0.10) & 0.265 (0.024) \\
938   & 150  & 19.02 (0.16) & 0.090 (0.014) & 18.60 (0.20) & 0.132 (0.025) & 17.92 (0.10) & 0.246 (0.023) \\
1163  & 300  & 19.35 (0.16) & 0.066 (0.010) & 18.84 (0.11) & 0.106 (0.011) & 18.00 (0.09) & 0.229 (0.019) \\
1463  & 300  & 19.62 (0.18) & 0.051 (0.009) & 19.12 (0.19) & 0.081 (0.014) & 18.27 (0.11) & 0.179 (0.017) \\
2304  & 600  & 19.37 (0.09) & 0.065 (0.006) & 19.36 (0.10) & 0.066 (0.006) & 18.69 (0.13) & 0.122 (0.014) \\
2904  & 600  & 19.83 (0.14) & 0.042 (0.005) & 19.22 (0.09) & 0.074 (0.006) & 18.54 (0.10) & 0.139 (0.013) \\
3504  & 600  & 19.57 (0.10) & 0.054 (0.005) & 19.14 (0.08) & 0.080 (0.006) & 18.52 (0.10) & 0.142 (0.013) \\
4104  & 600  & 19.21 (0.07) & 0.075 (0.005) & 19.07 (0.08) & 0.085 (0.006) & 18.52 (0.08) & 0.142 (0.010) \\
4704  & 600  & 19.57 (0.10) & 0.054 (0.005) & 19.29 (0.10) & 0.070 (0.006) & 18.48 (0.06) & 0.147 (0.008) \\
5640  & 600  & 19.38 (0.09) & 0.064 (0.005) & 19.28 (0.09) & 0.070 (0.006) & 18.46 (0.06) & 0.151 (0.008) \\
6420  & 600  & 19.20 (0.09) & 0.076 (0.006) & 19.44 (0.11) & 0.061 (0.006) & 18.32 (0.13) & 0.171 (0.021) \\
\hline
\end{tabular}
\end{adjustbox}

\vspace{2mm}
\parbox{\textwidth}{\footnotesize \textit{Note.}Calibration procedures involved referencing 4 bright stars from the Pan-STARRS catalog within our images with object IDs:
204052911191064516,
203992906791944462,
204042913023972808 and
204042912675745804.
}

\end{table*}

\begin{table*}
\centering
\scriptsize
\setlength{\tabcolsep}{3pt}
\caption{Later observations. $t_\mathrm{mid}$ is the midpoint time of each co-add, and $t_0$ is the BAT trigger time, which occurred at 14:54:48 UT on 28 March 2023 \citep{Gropp2023GCN}. $T_\mathrm{exp}$ denotes the exposure time of each image. The Milky Way extinction is subtracted.\label{tab:later_obs}}
\begin{tabular}{cccccccc}
\hline
Date & Site/Instrument & Band & $t_\mathrm{mid}-t_0$ [s] & $T_\mathrm{exp}$ [s] & mag (err) & Flux (err) [mJy] & References \\
\hline

2023-03-28 & Mondy/AZT-33IK & $R$ & 2251 & 600.0 & 19.38 (0.07) & 0.064 (0.004) & This work\\
2023-03-28 & Mondy/AZT-33IK & $R$ & 2852 & 600.0 & 19.45 (0.08) & 0.060 (0.005) & This work\\
2023-03-28 & Mondy/AZT-33IK & $R$ & 3452 & 600.0 & 19.29 (0.09) & 0.070 (0.006) & This work\\
2023-03-28 & Mondy/AZT-33IK & $R$ & 4053 & 600.0 & 19.39 (0.10) & 0.064 (0.006) & This work\\
2023-03-28 & Mondy/AZT-33IK & $R$ & 4653 & 600.0 & 19.38 (0.09) & 0.064 (0.006) & This work\\
2023-03-28 & Mondy/AZT-33IK & $R$ & 5373 & 600.0 & 19.38 (0.07) & 0.064 (0.004) & This work\\
2023-03-28 & Mondy/AZT-33IK & $R$ & 5975 & 600.0 & 19.48 (0.06) & 0.058 (0.003) & This work\\
2023-03-28 & Mondy/AZT-33IK & $R$ & 6575 & 600.0 & 19.45 (0.06) & 0.060 (0.004) & This work\\
2023-03-28 & Mondy/AZT-33IK & $R$ & 7176 & 600.0 & 19.50 (0.06) & 0.058 (0.004) & This work\\
2023-03-28 & Mondy/AZT-33IK & $R$ & 7776 & 600.0 & 19.45 (0.06) & 0.060 (0.004) & This work\\
2023-03-28 & Mondy/AZT-33IK & $R$ & 8377 & 600.0 & 19.61 (0.06) & 0.052 (0.003) & This work\\
2023-03-28 & Mondy/AZT-33IK & $R$ & 8977 & 600.0 & 19.83 (0.07) & 0.043 (0.003) & This work\\
2023-03-28 & Mondy/AZT-33IK & $R$ & 18521 & 3480.0 & 20.33 (0.09) & 0.026 (0.002) & This work\\
2023-03-29 & Mondy/AZT-33IK & $R$ & 88340 & 3600 & 21.81 (0.09) & 0.007 (0.001) & This work\\
2023-04-05 & Mondy/AZT-33IK & $R$ & 693297 & 7200 & $>22.5$ & $<0.004$ & This work\\
2023-04-11 & Mondy/AZT-33IK & $R$ & 1225109 & 7200 & 23.07 (0.21) & 0.002 (0.001) & This work\\
2023-04-13 & Mondy/AZT-33IK & $R$ & 1395378 & 7080 & 23.43 (0.58) & 0.002 (0.001) & This work\\
2023-04-14 & Mondy/AZT-33IK & $R$ & 1472671 & 8400 & 23.03 (0.20) & 0.002 (0.001) & This work\\
2023-04-22 & Mondy/AZT-33IK & $R$ & 2213162 & 10800 & 23.05 (0.11) & 0.002 (0.001) & This work\\
2023-03-28 & ISON-Multa/Santel-400 & $Clear$ & 3977 & 5400 & 18.87 (0.13) & 0.103 (0.011) & This work\\
2023-03-28 & ISON-Kitab/RC-36 & $Clear$ & 3658 & 1140 & $>17.6$ & $<0.330$ & This work\\
2023-03-28 & IAO/GIT & $r$ & 8638 & 1800 & 19.30 (0.13) & 0.068 (0.008) & This work \\
2023-03-28 & IAO/GIT & $r$ & 13064 & 1500 & 19.93 (0.15) & 0.038 (0.005) & This work \\
2023-03-28 & IAO/GIT & $r$ & 25926 & 600 & 20.23 (0.18) & 0.029 (0.005) & This work\\
2023-03-28 & IAO/GIT & $g$ & 8636 & 1200 & 19.64 (0.12) & 0.050 (0.005) & This work\\
2023-03-28 & IAO/GIT & $g$ & 25254 & 600 & 21.01 (0.35) & 0.014 (0.004) & This work\\
2023-03-28 & IAO/GIT & $i$ & 9954 & 1200 & 19.49 (0.23) & 0.057 (0.012) & This work\\
2023-03-28 & IAO/GIT & $i$ & 26581 & 600 & 19.96 (0.20) & 0.037 (0.007) & This work\\
2023-03-28 & AbAO/AS-32 & $R$ & 7493 & 6240 & 19.48 (0.10) & 0.059 (0.006) & This work\\
2023-03-28 &  UAFO/RC-500 & $Clear$ & 8408 & 4050 & 19.32 (0.15) & 0.068 (0.009) & This work\\
2023-03-28 & ARIES/DOT & $R$ & 25026 & 300 & 20.46 (0.03) & 0.024 (0.001) & This work\\
2023-03-28 & ARIES/DOT & $R$ & 25566 & 300 & 20.57 (0.04) & 0.021 (0.001) & This work\\
2023-03-28 & ARIES/DOT & $R$ & 25876 & 300 & 20.48 (0.03) & 0.023 (0.001) & This work\\
2023-03-28 & ARIES/DOT & $R$ & 28423 & 300 & 20.56 (0.03) & 0.022 (0.001) & This work\\
2023-03-28 & ARIES/DOT & $R$ & 28734 & 300 & 20.57 (0.05) & 0.021 (0.001) & This work\\
2023-03-28 & ARIES/DOT & $R$ & 29044 & 300 & 20.61 (0.04) & 0.021 (0.001) & This work\\
2023-03-29 & ARIES/DOT & $R$ & 107600 & 300 & 22.04 (0.05) & 0.005 (0.001) & This work\\
2023-04-15 & ORM/GTC & $r$ & 1518033 & 60 & 23.70 (0.25) & 0.001 (0.000) & This work\\
2023-04-15 & SAO/BTA & $g$ & 1581153 & 600 & 23.56 (0.13) & 0.001 (0.001) & This work\\
2023-05-15 & SAO/BTA & $g$ & 4184745 & 2280 & 23.37 (0.04) & 0.001 (0.001) & This work\\
2023-04-15 & SAO/BTA & $r$ & 1583148 & 700 & 23.09 (0.10) & 0.002 (0.001) & This work\\
2023-05-14 & SAO/BTA & $r$ & 4094954 & 2400 & 23.02 (0.04) & 0.002 (0.001) & This work\\
2023-04-15 & SAO/BTA & $i$ & 1581129 & 1200 & 22.85 (0.10) & 0.003 (0.001) & This work\\
2023-05-15 & SAO/BTA & $i$ & 4188717 & 3960 & 22.88 (0.04) & 0.003 (0.001) & This work\\
2023-04-23 & CrAO/ZTSh & $R$ & 2275085 & 7440 & 23.47 (0.27) & 0.001 (0.001) & This work\\
2023-05-13 & MAO/AZT-22 & $I$ & 4000776 & 3600 & 22.85 (0.27) & 0.001 (0.001) & This work\\
2023-05-19 & MAO/AZT-22 & $R$ & 4517448 & 3600 & 23.19 (0.35) & 0.002 (0.001) & This work\\
\hline
2023-03-30 & CrAO/RT-22 & 36.8 GHz & 152172 & 12180 & - & 2.3 (1.2) & This work\\
2023-03-31 & CrAO/RT-22 & 36.8 GHz & 239172 & 9360 & - & 2.8 (1.2) & This work\\
2023-04-01 & CrAO/RT-22 & 36.8 GHz & 329172 & 11400 & - & 3.4 (1.3) & This work\\
2023-04-02 & CrAO/RT-22 & 36.8 GHz & 418032 & 9960 & - & 3.3 (1.1) & This work\\
2023-04-03 & CrAO/RT-22 & 36.8 GHz & 505692 & 9000 & - & 3.2 (1.1) & This work\\
2023-04-04 & CrAO/RT-22 & 36.8 GHz & 591372 & 5820 & - & 2.6 (1.2) & This work\\
2023-04-05 & CrAO/RT-22 & 36.8 GHz & 680772 & 5820 & - & 1.8 (1.2) & This work\\
\hline
2023-03-28 & \textit{Swift}/UVOT & $white$ & 188 & 147 & - & 0.075 (0.003) & This work\\           
2023-03-28 & \textit{Swift}/UVOT & $white$ & 5635 & 196 & - & 0.015 (0.001) & This work\\          
2023-03-28 & \textit{Swift}/UVOT & $u$ & 451 & 246 & - & 0.034 (0.005) & This work\\               
2023-03-28 & \textit{Swift}/UVOT & $u$ & 5226 & 246 & - & 0.015 (0.004) & This work\\            
2023-03-28 & \textit{Swift}/UVOT & $v$ & 99 & 246 & - & 0.322 (0.113) & This work\\                
2023-03-28 & \textit{Swift}/UVOT & $v$ & 4611 & 246 & - & 0.040 (0.016) & This work\\              
2023-03-28 & \textit{Swift}/UVOT & $b$ & 5431 & 246 & - & 0.029 (0.007) & This work\\   
\hline
2023-03-28 & OHP/T120 & $R_\mathrm{c}$ & 16200 & 3000 & 20.08 (0.10) & 0.033 (0.003) & \cite{Adami2023} \\
2023-03-28 & GRANDMA/KNC & $R_\mathrm{c}$ & 19157 & 2160 & 20.24 (0.10) & 0.029 (0.003) & \cite{Kugel2023} \\
2023-03-28 & CAHA/CAFOS & $R_\mathrm{c}$ & 40754 & 1800 & 20.84 (0.06) & 0.016 (0.003) & \cite{Agui2023} \\
2023-03-28 & ORM/Liverpool & $i'$ & 44640 & 1800 & 20.63 (0.07) & 0.020 (0.001) & \cite{Gompertz2023} \\
2023-03-28 & ORM/Liverpool & $r'$ & 44640 & 1800 & 21.00 (0.06) & 0.014 (0.001) &  \cite{Gompertz2023} \\
\hline
\end{tabular}
\end{table*}

\begin{figure*}
    \centering
    \includegraphics[width=1.0\linewidth]{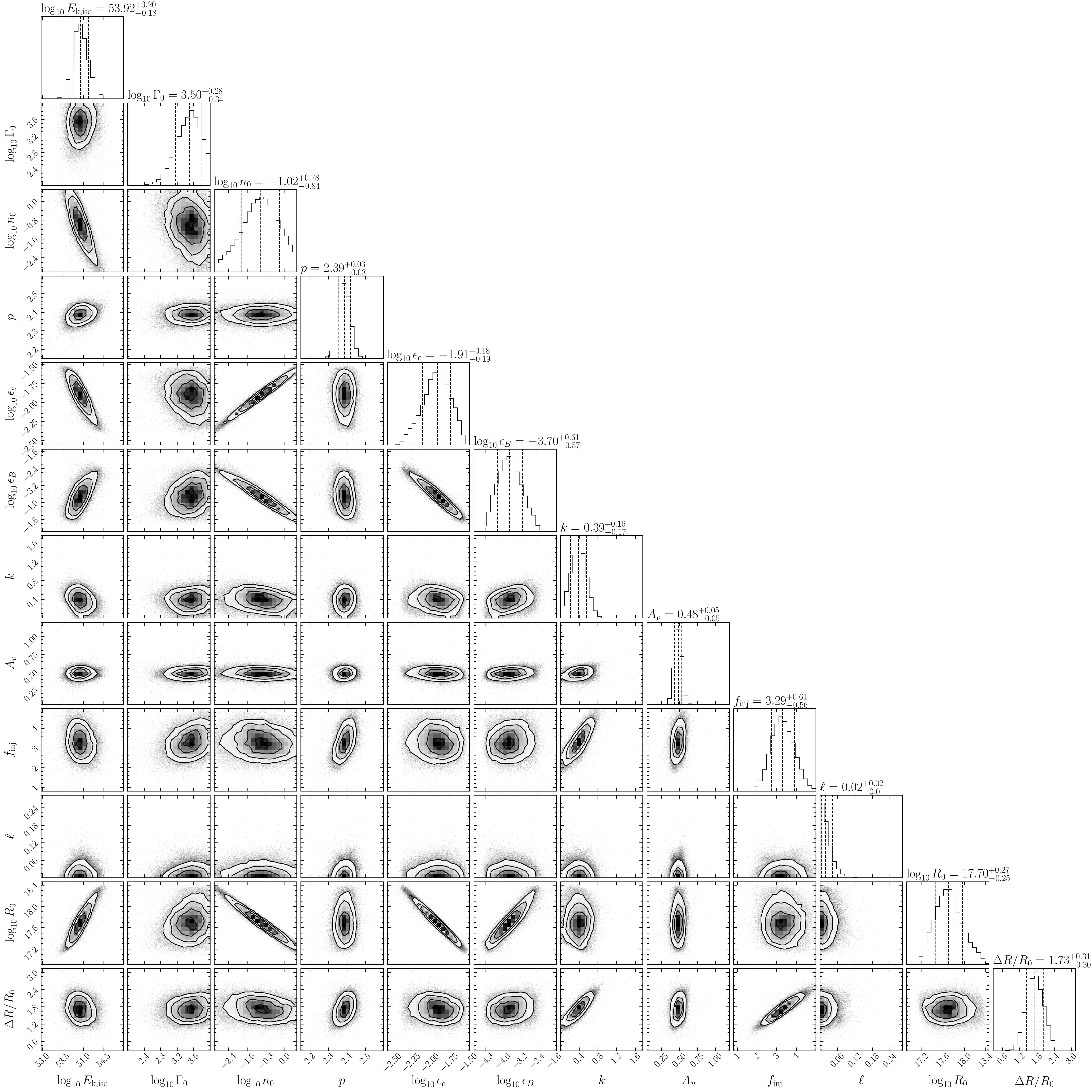}
    \caption{Corner plot of the posterior parameter distributions from the MCMC analysis. The diagonal panels show marginalized one-dimensional distributions, with vertical lines indicating the 16th, 50th (median), and 84th percentiles (corresponding to the $1\sigma$ credible intervals). The off-diagonal panels display two-dimensional joint constraints with density contours. }
    \label{fig:cp_MCMC}
\end{figure*}

%%%%%%%%%%%%%%%%%%%%%%%%%%%%%%%%%%%%%%%%%%%%%%%%%%

% Don't change these lines
\bsp	% typesetting comment
\label{lastpage}
\end{document}